\pdfoutput=1
\documentclass[11pt,twoside,a4paper,cmspaper,final,collab]{cms-tdr}
\def\svnVersion{1fbd416}\def\svnDate{2026/07/10}\def\cmsCernNoTag{CERN-EP-2025-228}\def\cmsCernDate{\today}\def\cmsMessage{Published in the Journal of High Energy Physics as \href{http://dx.doi.org/10.1007/JHEP07(2026)041}{\doi{10.1007/JHEP07(2026)041}.}}
\begin{document}\cmsNoteHeader{B2G-23-009}

\newcommand{\wjets}{\PW{+}\text{jets}\xspace}
\newcommand{\zjet}{\PZ{+}\text{jet}\xspace}
\newcommand{\DeepJet}{\textsc{DeepJet}\xspace}
\newcommand{\MlvbJ}{\ensuremath{M_{\Pell\PGn\PQb\mathrm{J}}}\xspace}
\newcommand{\Ptrue}{\ensuremath{\PQt_{\text{true}}}\xspace}
\newcommand{\PQCD}{\ensuremath{\PQt_{\text{QCD}}}\xspace}
\newcommand{\POth}{\ensuremath{\PQt_{\text{oth}}}\xspace}
\newcommand{\Imini}{\ensuremath{I_{\text{mini}}(\Pell)}\xspace}
\newcommand{\Xbb}{\ensuremath{\PX_{\PQb\PQb}}\xspace}
\newcommand{\MXbb}{\ensuremath{M_{\Xbb}}\xspace}
\newcommand{\ptrell}{\ensuremath{\pt^\text{rel}(\Pell)}\xspace}
\newcommand{\ttH}{\ensuremath{\ttbar\PH}\xspace}

\cmsNoteHeader{B2G-23-009}

\title{Search for single production of a vector-like \texorpdfstring{\PQT}{T} quark decaying to a top quark and a neutral scalar boson in the lepton+jets final state in proton-proton collisions at \texorpdfstring{$\sqrt{s} = 13\TeV$}{sqrt(s) = 13 TeV}}

\date{\today}

\newlength\cmsTabSkip\setlength{\cmsTabSkip}{1ex}

\abstract{A search for single production of a vector-like \PQT quark with charge $2e/3$, decaying to a top quark and a neutral scalar boson is presented. The boson can be a standard model Higgs boson (\PH) or a new scalar boson ($\PGf$). In the first case, a branching fraction of 25\% is assumed for the decay $\PQT\to\PQt \PH$, while in the second case the \PQT quark is assumed to decay exclusively to $\PQt \PGf$. The top quark is identified via its lepton+jets decay, and the neutral boson via its decay into a bottom quark-antiquark pair. Final states with Lorentz-boosted topologies are considered and machine-learning techniques are exploited for optimal event classification. The analysis uses data collected by the CMS experiment in proton-proton collisions at a center-of-mass energy of 13\TeV, corresponding to an integrated luminosity of 138\fbinv recorded at the CERN LHC in 2016--2018. Upper limits at 95\% confidence level are set on the product of cross section and branching fraction for a \PQT quark in a narrow-width approximation. They vary between 14.7 and 0.1\unit{fb}, for \PQT quark masses in the range 1.3--3.0\TeV and $\PGf$ boson masses in the range 25--250\GeV. These are the first exclusion limits set on the production of a single \PQT quark decaying into a top quark and a new neutral scalar boson. For the decay channel into a top quark and a standard model Higgs boson, the results provide the best limits on production cross sections to date, for \PQT quark masses above 2\TeV.}

\hypersetup{
pdfauthor={CMS Collaboration},
pdftitle={Search for single production of a vector-like T quark decaying to a top quark and a neutral scalar boson in the lepton+jets final state in proton-proton collisions at sqrt(s) = 13 TeV},
pdfsubject={CMS},
pdfkeywords={CMS, vector-like quarks}}

\maketitle 

\section{Introduction}\label{sec:intro}
Several extensions of the standard model (SM) predict the existence of heavy partners of SM quarks, referred to as vector-like
quarks (VLQs)~\cite{Aguilar-Saavedra:2013qpa, Okada:2012gy, AguilarSaavedra:2009es, DeSimone:2012fs, Matsedonskyi:2014mna, Buchkremer:2013bha, Carvalho:2018}, which provide solutions to unresolved issues of the theory such as the large gap between the Planck scale and the electroweak scale, known as the hierarchy problem. These VLQs are colored spin-$1/2$ fermions with symmetric chirality under the electroweak gauge symmetry group. Therefore, unlike the chiral quarks of the SM, VLQs do not receive their masses from Yukawa couplings.
While precision SM measurements have put tight constraints on the existence of a fourth generation of chiral quarks~\cite{Eberhardt:2012sb}, this is not the case for VLQs.
At the CERN LHC, VLQs may be produced in pairs mainly via strong processes, or singly via electroweak processes.
The latter mode dominates for VLQ masses above 1\TeV owing to the larger phase space.

\begin{figure}[htbp]
\centering
 \includegraphics[width=0.58\textwidth]{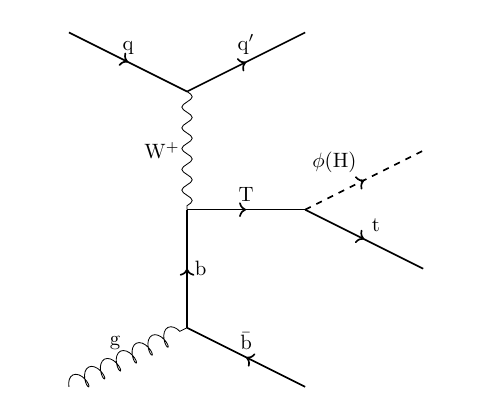}
 \caption{A representative leading-order Feynman diagram for the production of a single vector-like quark \PQT decaying into a top quark
   and a neutral scalar boson, which may be a new scalar boson or an SM Higgs boson. Associated production with a bottom quark in the final state is shown.}
\label{fig:feynman}
\end{figure}

This paper presents a search for the production of a single vector-like \PQT quark with charge $2e/3$ in proton-proton ($\Pp\Pp$) collision data at $\sqrt{s}=13\TeV$ collected with the CMS detector in 2016--2018 and corresponding to an integrated luminosity of 138\fbinv.
The decay channels analyzed are $\PQT\to\PQt \PGf$ and $\PQT\to\PQt \PH$ where \PQt is a top quark, $\PGf$ is a new neutral scalar boson, and \PH is the SM Higgs boson.
An example of a leading-order (LO) Feynman diagram for single \PQT quark production,
where the \PQT quark is produced in association with a bottom (\PQb) quark in the final state ($\PQb\PQT\PQq$ production)
is shown in Fig.~\ref{fig:feynman}, with the \PQT quark decaying to $\PQt \PGf$ or $\PQt \PH$.

In minimal scenarios, which have been typically considered for VLQ searches, the \PQT quark couples to third-generation SM quarks and weak bosons,
leading to decay channels into a bottom quark and a \PW boson, or a top quark and a \PZ boson. It also couples to the Higgs boson, resulting in the $\PQt\PH$ decay. Branching fractions are a function of the model-dependent values of the couplings.
In this search, we assume a singlet \PQT quark, which decays into $\PQt\PH$ with a branching fraction of 25\%~\cite{Aguilar-Saavedra:2013qpa, CMS:2024bni}.
A nonminimal scenario, where the \PQT quark decays to a top quark and a beyond-the-SM scalar boson
with a 100\% branching fraction~\cite{Aguilar-Saavedra:2019ghg,Benbrik:2019zdp,Cacciapaglia:2019zmj},
is also considered in this search.

Recent reviews of VLQ searches in Run 2 of the LHC have been published by the ATLAS~\cite{ATLAS:2024fdw} and CMS~\cite{CMS:2024bni} Collaborations.
Searches for pair production of \PQT quarks at a center-of-mass energy of 13\TeV, targeting a variety of possible final states, have been performed by ATLAS~\cite{Aaboud:2017qpr, Aaboud:2017zfn, Aaboud:2018saj, Aaboud:2018wxv, Aaboud:2018xuw, Aaboud:2018pii, ATLAS:2022tla, ATLAS:2022hnn, ATLAS:2024gyc, ATLAS:2024zlo} and CMS~\cite{Sirunyan:2017usq, Sirunyan:2017pks, Sirunyan:2018omb, Sirunyan:2018qau, Sirunyan:2019sza, CMS:2022fck}.
Searches for single production of \PQT quarks in the $\PQt\PH$ decay channel, exploiting fully hadronic final states, have also been reported by ATLAS~\cite{ATLAS:2022ozf} and CMS~\cite{Sirunyan:2019xeh, CMS:2024qdd}. A search in the $\PQt\PH$ and $\PQt\PZ$ channels using final states with one lepton has been published by ATLAS~\cite{ATLAS:2023pja}, while a search exploiting the diphoton decay mode of the Higgs boson has been published by CMS~\cite{CMS:2023agg}. The $\PQt\PZ$ and $\PQb\PW$ decay channels have also been investigated, with results reported by both ATLAS~\cite{Aaboud:2018zpr, Aaboud:2018ifs, ATLAS:2023bfh, ATLAS:2024xne, ATLAS:2024kgp} and CMS~\cite{Sirunyan:2017ynj, CMS:2022yxp, Sirunyan:2017tfc}. 

The search reported in this paper targets final states where the top quark decays in the lepton+jets channel
via $\PQt \to \PW\PQb \to \Pell\nu\PQb$ and the scalar boson decays to a bottom quark-antiquark
pair ($\PH, \PGf\to\bbbar$).
The experimental signature consists of one electron or muon
(denoted as $\Pell$ in this paper),
the hadronization products of the \PQb quarks,
and a neutrino that is not reconstructed in the detector,
causing an imbalance in the total transverse momentum (\pt) of the reconstructed events.
As shown in Fig.~\ref{fig:feynman}, in addition to the bottom quark produced in association with
the \PQT quark, final states of signal events are also characterized by a light-flavor quark that tends to be emitted
at a low angle with respect to the beam axis, in the forward region of the detector.
Mass values in the 1--3\TeV range for the \PQT quark,
and 25--250\GeV for the $\PGf$ boson, are considered in the analysis.

The analysis is designed to be sensitive to
signatures with a large Lorentz boost of the \PQT quark decay products.
The $\PGf$ boson is reconstructed as a large-radius jet,
since in this regime the \PQb quarks from its decay
produce jets that tend to be collimated.
Similarly, the lepton and jet from
the top quark decay tend to overlap,
which makes the top quark reconstruction more challenging.
A top quark identification algorithm, based on machine-learning techniques,
has been developed for this analysis in order to optimize the identification efficiency
for the top quark decay.

For the interpretation of search, a singlet \PQT quark model~\cite{Matsedonskyi:2014mna}
is assumed, in which the production process is associated $\PQb\PQT\PQq$ production.
This process has a much larger cross section than the associated production with a top quark $\PQt\PQT\PQq$,
which is in fact suppressed because of the higher masses of top quark as compared to bottom quark~\cite{Matsedonskyi:2014mna},
and is therefore not considered as part of the signal.
However, no specific requirement on the associated top or bottom quark is considered
in the analysis selection and strategy, and, by design, the results of this search may also be interpreted
in terms of other models, such as composite Higgs or little Higgs models~\cite{Buchkremer:2013bha}, with the same final state.

Tabulated results are provided in the HEPData record for this analysis~\cite{hepdata}.

\section{The CMS detector, data, and simulation}\label{sec:detector}
The central feature of the CMS apparatus is a superconducting solenoid of 6\unit{m} internal diameter, providing a magnetic field of 3.8\unit{T}. Within the solenoid volume are a silicon pixel and strip tracker, a lead tungstate crystal electromagnetic calorimeter (ECAL), and a brass and scintillator hadron calorimeter (HCAL), each composed of a barrel and two endcap sections. Forward calorimeters extend the pseudorapidity coverage provided by the barrel and endcap detectors. Muons are measured in gas-ionization detectors interleaved with the layers of the steel flux-return yoke outside the solenoid. More detailed descriptions of the CMS detector, together with a definition of the coordinate system used and the relevant kinematic variables, can be found in Refs.~\cite{Chatrchyan:2008zzk, CMS:2023gfb}.

Events of interest are selected using a two-tiered trigger system. The first level (\Lone), composed of custom hardware processors, uses information from the calorimeters and muon detectors to select events at a rate of around 100\unit{kHz} within a fixed latency of 4\mus~\cite{CMS:2020cmk}. The second level, known as the high-level trigger, consists of a farm of processors running a version of the full event reconstruction software optimized for fast processing, and reduces the event rate to a few \unit{kHz} before data storage~\cite{CMS:2016ngn,CMS:2024aqx}.

Signal events of this analysis are selected online by triggers that require a muon, an electron, a photon, or a large-radius jet to be present in the final state.
The photon requirement is added to recover efficiency for events with a very high \pt electron.
Different trigger thresholds are used in the different data-taking periods because of changes in the instantaneous luminosity and detector configuration.
The high-level trigger selection requires an event to satisfy at least one of the following criteria:
presence of a muon with $\pt > 50\GeV$, or an isolated muon with $\pt > 24$~$(27)\GeV$ for 2016 and 2018 (2017) data;
presence of an isolated electron with $\pt > 27$\,(32)\GeV for 2016 (2017--2018) data;
presence of a photon with $\pt > 175$\,(200)\GeV for 2016 (2017--2018) data;
presence of a large-radius jet with mass ${>}30\GeV$ and $\pt > 360,$ 500 and 400\GeV for 2016, 2017 and 2018 data, respectively.

The combination of the above trigger requirements has an efficiency above 90\% for signal events, estimated from simulated samples whose details are described in the following.

Simulated background samples are generated at LO with \MGvATNLO \cite{Alwall:2014hca}, version 2.6.5, for \wjets events and for events with jets arising exclusively from quantum chromodynamics (QCD) interactions, henceforth referred to as QCD multijet events.
The next-to-LO (NLO) generator \POWHEG 2.0 \cite{Nason:2004rx, Frixione:2007vw, Alioli:2010xd} is used to simulate top quark-antiquark pair (\ttbar) events~\cite{Frixione:2007nw}, production of \ttbar events in association with a Higgs boson (\ttH)~\cite{Hartanto:2015uka}, and single top quark events produced in the $t$~channel~\cite{fourfiveflavorschemes} or in association with a \PW boson~\cite{Re:2010bp}.
Single top quark events produced in the $s$~channel are generated at NLO with \MGvATNLO; this generator and the \POWHEG are interfaced with \PYTHIA 8.240~\cite{Sjostrand:2014zea} for hadronization and parton shower simulation, with the CP5 underlying event tune~\cite{Sirunyan:2019dfx}.
The background processes are initially normalized to their theoretical cross sections, using the highest order available.
The cross section for the \ttbar background is computed at next-to-NLO (NNLO) in perturbative QCD using a soft-gluon resummation at next-to-next-to-leading logarithmic precision with the \textsc{Top++} 2.0 program \cite{Czakon:2011xx}. 
For single top quark production, NLO calculations are performed with \textsc{hathor} 2.1~\cite{hator}.
The \wjets cross section is calculated at LO and corrected with an inclusive $K$ factor to account
for NLO electroweak and NNLO QCD corrections~\cite{Lindert:DYCorrections}.

Signal event samples are generated at LO using \MGvATNLO interfaced to \PYTHIA.
We assume a singlet \PQT quark that is produced in association with a bottom quark with suppressed right-handed coupling to SM particles~\cite{Cacciapaglia:2011fx}.
The generated \PQT quark mass in the samples ranges from 1 to 3\TeV, in steps of 0.1\TeV, and the \PQT quark width $\Gamma$ is set to 10\GeV, a value much smaller than the experimental resolution, which is of the order of 200--300\GeV. The generated mass of the $\PGf$ boson ranges from 25 to 200\GeV, in steps of 25\GeV. An additional sample is generated with a mass of 250\GeV.
All possible leptonic decays of the top quark are simulated for the signal process. 

For the interpretation of the results in the case of the $\PQt\PH$ decay channel, the total cross section is computed at NLO
in the narrow-width approximation using the simplified approach from Refs.~\cite{Buchkremer:2013bha, Carvalho:2018}, for a singlet \PQT quark assuming a resonance width $\Gamma$ of 5 or 1\% of the mass.

The simulated samples are generated using the NNLO NNPDF~3.1~\cite{Ball:2014uwa} parton distribution function (PDF) sets.

{\tolerance=1500 The generated events are passed through a simulation of the CMS detector based on \GEANTfour~\cite{Agostinelli:2002hh, Allison:2006ve}.
Additional interactions in the same or adjacent bunch crossings (pileup) are included in the simulation, with their multiplicity distribution adjusted to match that observed in data.\par}

\section{Event reconstruction and selection}\label{sec:reconstruction}
The primary vertex (PV) is taken to be the vertex corresponding to the hardest scattering in the event, evaluated using tracking information alone, as described in Section 9.4.1 of Ref.~\cite{CMS-TDR-15-02}.
It must be within 24\unit{cm} of the nominal interaction point along the beam axis and within 
2\unit{cm} in the transverse plane.

A particle-flow (PF) algorithm~\cite{CMS:2017yfk} aims to reconstruct and identify each individual particle in an event, with an optimized combination of information from the various elements of the CMS detector. The energy of photons is obtained from the ECAL measurement. The energy of electrons is determined from a combination of the electron momentum at the PV as determined by the tracker, the energy of the corresponding ECAL cluster, and the energy sum of all bremsstrahlung photons spatially compatible with originating from the electron track. The energy of muons is obtained from the curvature of the corresponding track. The energy of charged hadrons is determined from a combination of their momentum measured in the tracker and the matching ECAL and HCAL energy deposits, corrected for the response function of the calorimeters to hadronic showers. Finally, the energy of neutral hadrons is obtained from the corresponding corrected ECAL and HCAL energies.

Jets are clustered from the PF candidates using the anti-\kt algorithm~\cite{Cacciari:2008gp, Cacciari:2011ma} with 
distance parameters of 0.4 (AK4 jets) and 0.8 (AK8 jets).
Jet momentum is determined as the vectorial sum of all particle momenta in the jet, and is found from simulation to be, on average, within 5--10\% of the true momentum over the entire spectrum and detector acceptance. Pileup interactions can contribute additional tracks and calorimetric energy depositions, increasing the apparent jet momentum.
To mitigate this effect for the AK4 jets, tracks identified to be originating from pileup vertices are discarded and an offset correction is applied to correct for remaining contributions~\cite{CMS:2020ebo}.
For the AK8 jets the pileup-per-particle identification algorithm~\cite{CMS:2020ebo, Bertolini:2014bba} is used to mitigate the effect of pileup at the reconstructed-particle level, making use of local shape information, event pileup properties, and tracking information. A local shape variable is defined, which distinguishes between collinear and soft diffuse distributions of other particles surrounding the particle under consideration. The former is attributed to particles originating from the hard scatter and the latter to particles originating from pileup interactions. Charged particles identified to be originating from pileup vertices are discarded. For each neutral particle, a local shape variable is computed using the surrounding charged particles compatible with the PV within the tracker acceptance ($\abs{\eta} < 2.5$), and using both charged and neutral particles in the region outside of the tracker coverage. The momenta of the neutral particles are then rescaled according to their probability to originate from the PV, deduced from the local shape variable, superseding the need for jet-based pileup corrections~\cite{CMS:2020ebo}.
Jet energy corrections are derived from simulation studies so that the average measured energy of jets becomes identical to that of particle level jets. In situ measurements of the momentum balance in dijet, photon+jet, \zjet, and QCD multijet events are used to determine any residual differences between the jet energy scale in data and in simulation, and appropriate corrections are made~\cite{CMS:2016lmd}. Additional selection criteria are applied to each jet to remove jets dominated by instrumental effects or reconstruction failures~\cite{CMS:2020ebo}.

The AK4 jets selected for the analysis must have $\pt>30\GeV$. In order to remove jets originating from pileup,
jets with $30<\pt<50\GeV$ are required to pass the tight working point (WP) of a boosted decision tree (BDT)
discriminator~\cite{CMS:2020ebo}, which takes jet shapes and kinematic variables as inputs. 
Signal events are characterized by the presence of a forward jet.
At reconstruction level, forward jets are defined as
AK4 jets in the region $2.4 < \abs{\eta} < 4.0$.
Jets with larger values of $\abs{\eta}$ are not included in the count of forward jets, in order to reduce the contribution
from pileup interactions. 

Muon (electron) candidates selected for the analysis must have $\pt>30$\,(35)\GeV,
$\abs{\eta} < 2.4$\,(2.5), a transverse distance from the PV less than 0.02\,(0.05)\cm,
and satisfy loose identification criteria~\cite{CMS:2020uim, Sirunyan:2018fpa}. Electron candidates are excluded if they are in the ECAL barrel-endcap transition region $1.44 < \abs{\eta} < 1.57$.
Selected events must have only one lepton passing the above requirements, in order to reduce contamination from \ttbar
with two leptons in the final state.
Signal events are characterized by a prompt lepton 
from the top quark decay. In order to discriminate prompt leptons
from misidentified ones or leptons from hadronic decay chains,
a relative isolation variable $I(\Pell)$~\cite{CMS:2020uim, Sirunyan:2018fpa} is
defined as the scalar \pt sum of the hadrons (charged and neutral) and photons in a cone around the lepton direction
of size $\Delta R = \sqrt{\smash[b]{ (\Delta \eta)^2 + (\Delta \PGf)^2}}$,
where $\Delta \eta$ and $\Delta \PGf$ are the differences with respect to the lepton direction in $\eta$ and $\PGf$.
The cone size is 0.4 for muons and 0.3 for electrons.
The sum is then corrected for pileup effects and divided by the lepton \pt. 
A modified relative isolation variable referred to as mini-isolation \Imini,
originally suggested in Ref.~\cite{Rehermann:2010vq}, is also defined, which is
optimized for leptons produced in boosted top quark decays by using a \pt-dependent cone size.
No isolation requirements are applied on muon and electron candidates used as
input for top quark reconstruction; instead, the $I(\Pell)$ and \Imini variables 
are employed as input variables to train a top quark tagging algorithm, based on a BDT and described in Section~\ref{sec:phitopTagging},
to identify true top quark candidates.

The missing transverse momentum vector \ptvecmiss is defined as the negative vector \pt sum
of all the PF candidates in an event, and its magnitude is denoted as \ptmiss~\cite{CMS:2019ctu}.
The \ptvecmiss is modified to account for corrections to the energy scale of the reconstructed jets in the event.
Selected events must have $\ptmiss > 25\GeV$.

\section{Boson and top quark candidate reconstruction}\label{sec:phitopTagging}
For the reconstruction of the $\PGf$ boson candidates, AK8 jets with $\pt > 500\GeV$ and $\abs{\eta} < 2.4$ are used as input.
An AK8 jet is tagged as originating from a Lorentz-boosted heavy-boson decay
to a bottom quark-antiquark pair ($\Xbb$ jet) by employing \textsc{ParticleNet-MD}~\cite{ParNet},
a mass-decorrelated (MD) algorithm. It is based on a dynamic graph convolutional neural network,
for identifying two-body hadronic decays of highly boosted particles.
The algorithm provides probability-like scores for the AK8 jet to originate
from two-prong hadronic decays of the type $\PX \to \bbbar$, $\PX \to \ccbar$,
and $\PX \to \qqbar$, or to be a background jet from QCD multijet events. The use of an MD algorithm allows different hypotheses for the $\PGf$ boson mass to be probed without losing signal efficiency.
The \textsc{ParticleNet-MD} scores $p(\Xbb)$ and $p(\mathrm{QCD})$ are combined in the XbbVsQCD variable:
\begin{equation}
\mathrm{XbbVsQCD} = \frac{p(\Xbb)}{p(\Xbb)+p(\mathrm{QCD})},
\end{equation}
and a threshold on the XbbVsQCD variable is then employed to tag $\Xbb$ jets.

The mass of an AK8 jet is reconstructed by employing a mass regression technique
based on \textsc{ParticleNet-MD}.
Events are selected if they have at least one AK8 jet passing a loose WP
threshold of the XbbVsQCD tagger with a rejection rate for background jets of 90\% and an efficiency for signal jets of 70\%, evaluated on jets
from simulated QCD multijet and signal events, or if there is at least one AK8 jet with a mass between 60 and 220\GeV.

Top quark candidates are reconstructed from the combination
of one lepton candidate $\Pell$,
one AK4 jet with $\abs{\eta}<2.4$, and the \ptvecmiss of event,
following the algorithm described in Ref.~\cite{Sirunyan:2018rlu}. Here, the neutrino \pt is
assumed to be equal to \ptvecmiss and the longitudinal component of the neutrino momentum
is computed by requiring the invariant mass of the lepton-neutrino pair to be equal to the \PW boson mass~\cite{PDG2024}.
If more than one top quark candidate is reconstructed in the event, all of the candidates are retained for further processing.

In order to suppress combinatorial background, the distance between the lepton and jet directions in the $\eta$-$\PGf$ plane, $\Delta R_{\Pell j}$, must be less than 2.
If $\Delta R_{\Pell j} < 0.4$, the lepton is likely to overlap with the jet, and the top quark candidate
is said to be in the merged configuration. In this case, the lepton four-momentum is subtracted from the jet four-momentum
before computing the top quark candidate four-momentum.
Otherwise, if $0.4 < \Delta R_{\Pell j} < 2.0$, the top quark candidate
is said to be in the resolved configuration.
A deep neural network based tagger, {\DeepJet}~\cite{CMS-BTV-16-002, Bols:2020bkb, CMS-DP-2018-058}, is used to
identify AK4 jets stemming from the hadronization of \PQb quarks, by exploiting information from the tracks,
neutral particles, and the secondary vertices within the jet.
The output of the \PQb tagger is employed as one of the input variables of the top quark tagging algorithm,
based on a BDT, which is described below.

Simulated samples are used to train a machine learning algorithm to identify top quark decays. In these samples,
generator information at parton level is employed to perform an angular matching
of the top quark decay products with the leptons and jets reconstructed in the event.
A successful matching of both the lepton and the jet defines what we henceforth refer to as a true top quark candidate (\Ptrue).
The properties of \Ptrue candidates with respect to the combinatorial background have been studied.
Background candidates where neither the lepton nor the AK4 jet
originate from the decay of a top quark, which are typical of QCD multijet events
(\PQCD candidates), are distinguished from all other types of
background (\POth candidates).
The latter category includes, for instance, combinatorial candidates in genuine top quark events,
in which the lepton and the AK4 jet do not both originate from the decay
of the same top quark, which are typical of \ttbar background events.
A multiclass BDT algorithm, making use of the $\mathrm{XGBoost}$ library~\cite{xgboost}, is trained to distinguish \Ptrue from 
\PQCD and \POth candidates.
Separate BDTs are trained for the merged or resolved configuration of the top quark candidate,
for each lepton flavor (muon or electron), and for a top quark candidate with transverse
momentum \pt above 500\GeV (high-\pt candidate) or below (low-\pt candidate).

In total, eight different trainings have been performed, each using a different set of input variables,
which are related to the top quark candidate as well as to the associated lepton and AK4 jet.
The input variables are:
\begin{itemize}
\item the \pt of the jet ($\pt(j)$);
\item the mass of the jet ($M(j)$);
\item the value of the {\DeepJet} \PQb tagger score for the jet (\PQb~$\mathrm{tag}(j)$);
\item the \pt of the lepton ($\pt(\Pell)$);
\item the distance of closest approach of the lepton with respect to the PV in the transverse plane ($d_{xy}(\Pell)$) and its uncertainty ($\sigma_{xy}(\Pell)$);
\item the distance of closest approach of the lepton with respect to the PV along the longitudinal direction ($d_z(\Pell)$) and its uncertainty ($\sigma_z(\Pell)$);
\item the lepton isolation ($I(\Pell)$) and mini-isolation (\Imini);
\item the ratio of the lepton to jet \pt;
\item the separation $\Delta R_{\Pell j}$ between the lepton and jet directions in the $\eta$-$\PGf$ plane;
\item the component of the lepton momentum orthogonal to the jet axis (\ptrell);
\item the cosine of the angle between the lepton and the jet in the top quark candidate reference frame ($\cos{\theta}_{\Pell j}$);
\item the invariant mass of the lepton+jet system ($M(\Pell j)$);
\item the \pt of the lepton+jet system ($\pt(\Pell j)$);
\item the invariant mass of the top quark candidate ($M(\Pell \PGn j)$);
\item the transverse mass of the top quark candidate ($M_\mathrm{T}(\Pell \PGn j)$);
\item the \pt of the top quark candidate ($\pt(\Pell \PGn j)$).
\end{itemize}
The set of variables used in each of the trainings is reported in Table~\ref{tab:variables}.

For a given top quark candidate, the BDT classifier provides three probability-like scores:
$p(\Ptrue)$, $p(\PQCD)$, and $p(\POth)$.
Two variables, TopVsQCD and TopVsOther, are defined as follows in order to discriminate \Ptrue
candidates from \PQCD and \POth background candidates, respectively:
\begin{equation}
\mathrm{TopVsQCD} = \frac{p(\Ptrue)}{p(\Ptrue)+p(\PQCD)};
\end{equation}
\begin{equation}
\mathrm{TopVsOther} = \frac{p(\Ptrue)}{p(\Ptrue)+p(\POth)}.
\end{equation}
\begin{table}[!ht]
  \topcaption{List of the variables employed for each of the eight trainings of the multiclass BDT algorithm
    for top quark candidate identification. The trainings are different for each lepton channel, for the merged or resolved configuration of the top quark candidate, and for the \pt range of the top quark candidate with a threshold of 500\GeV.}
\label{tab:variables}
\centering
\resizebox{\columnwidth}{!}{
\begin{tabular}{ccccccccc}
& \multicolumn{4}{c}{Electron channel} & \multicolumn{4}{c}{Muon channel}  \\
& \multicolumn{2}{c}{Resolved} & \multicolumn{2}{c}{Merged} & \multicolumn{2}{c}{Resolved} & \multicolumn{2}{c}{Merged} \\
& Low-\pt & High-\pt & Low-\pt & High-\pt & Low-\pt & High-\pt & Low-\pt & High-\pt \\ \hline
$M(j)$             & \checkmark & \checkmark  & \checkmark & \checkmark & \checkmark & \checkmark & \checkmark & \checkmark \\
$\pt(j)$                    & \checkmark & \checkmark  & \checkmark &            & \checkmark &            &            &            \\
\PQb~$\mathrm{tag}(j)$          & \checkmark & \checkmark  & \checkmark & \checkmark & \checkmark & \checkmark & \checkmark & \checkmark \\
$\pt(\Pell)$                 & \checkmark &             &            &            &            &            &            &            \\
$d_{xy}(\Pell)$              & \checkmark & \checkmark  & \checkmark & \checkmark & \checkmark & \checkmark & \checkmark & \checkmark \\
$\sigma_{xy}(\Pell)$         & \checkmark &             &            &            & \checkmark &            &            &            \\
$d_z(\Pell)$                 & \checkmark & \checkmark  & \checkmark & \checkmark & \checkmark &            & \checkmark & \checkmark \\
$\sigma_z(\Pell)$            & \checkmark &             &            &            & \checkmark &            &            &            \\
$I(\Pell)$                   & \checkmark & \checkmark  & \checkmark & \checkmark & \checkmark & \checkmark & \checkmark & \checkmark \\
\Imini   & \checkmark & \checkmark  & \checkmark & \checkmark & \checkmark & \checkmark & \checkmark & \checkmark \\
$\pt(\Pell)/\pt(j)$          & \checkmark &             & \checkmark & \checkmark &            &            & \checkmark &            \\
$M(\Pell \PGn j)$   & \checkmark & \checkmark  & \checkmark & \checkmark & \checkmark & \checkmark & \checkmark & \checkmark \\
$\pt(\Pell \PGn j)$          & \checkmark &             & \checkmark &            & \checkmark &            & \checkmark & \checkmark \\
$M(\Pell j)$        & \checkmark & \checkmark  &            &            & \checkmark & \checkmark &            &            \\
$\pt(\Pell j)$          & \checkmark &             &            &            &            &            &            &            \\
$M_\mathrm{T}(\Pell \PGn j)$ & \checkmark & \checkmark  & \checkmark & \checkmark &            &\checkmark  &            & \checkmark \\
\ptrell    & \checkmark & \checkmark  & \checkmark & \checkmark &            & \checkmark & \checkmark & \checkmark \\
$\cos{\theta}_{\Pell j}$     & \checkmark &             & \checkmark &            & \checkmark &            &            & \checkmark \\
$\Delta R_{\Pell j}$         & \checkmark &             & \checkmark & \checkmark & \checkmark & \checkmark &            &            \\
\end{tabular}
}
\end{table}

\begin{figure}[h!t]
\centering
\includegraphics[width=.48\textwidth]{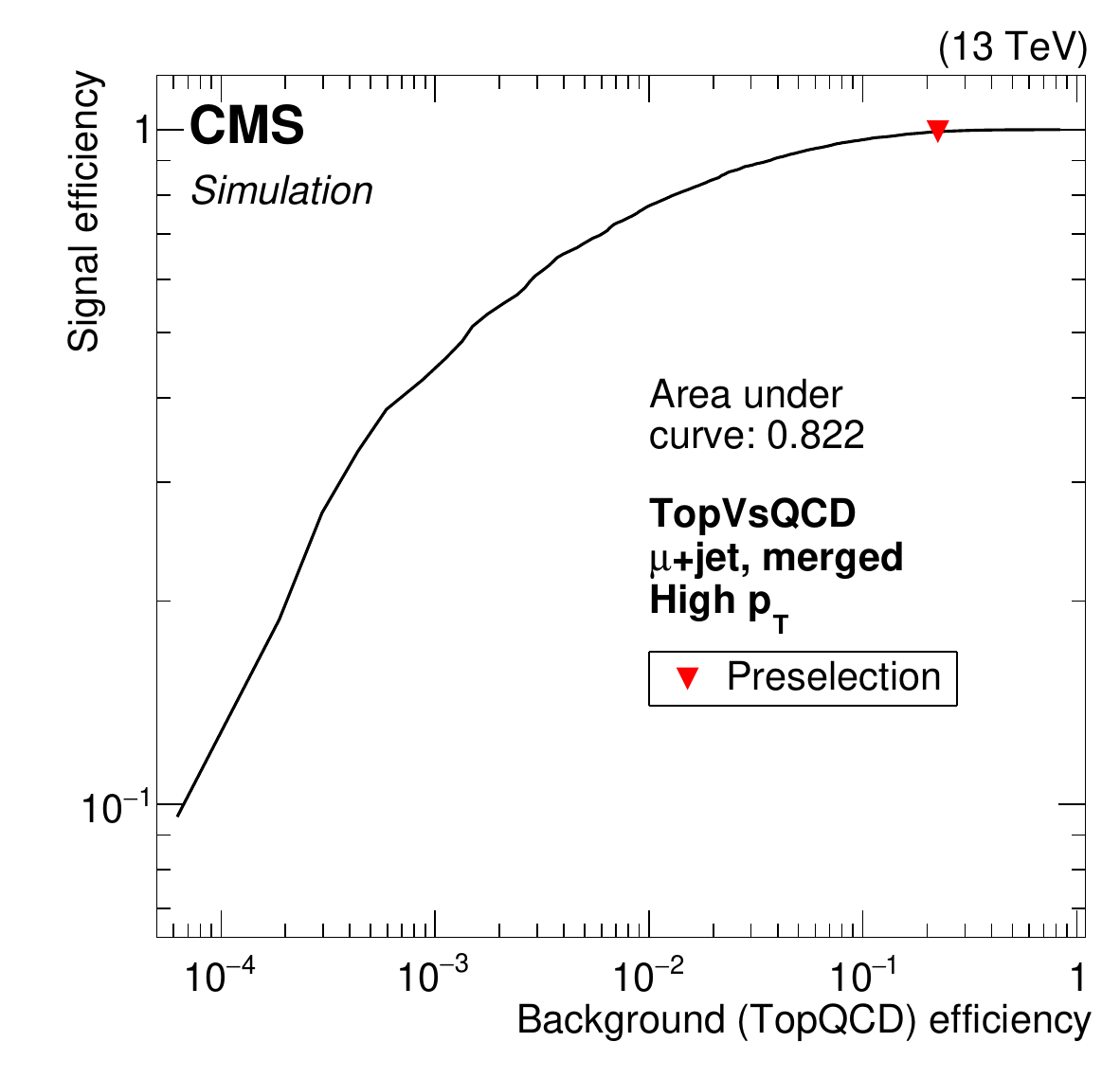}
\includegraphics[width=.48\textwidth]{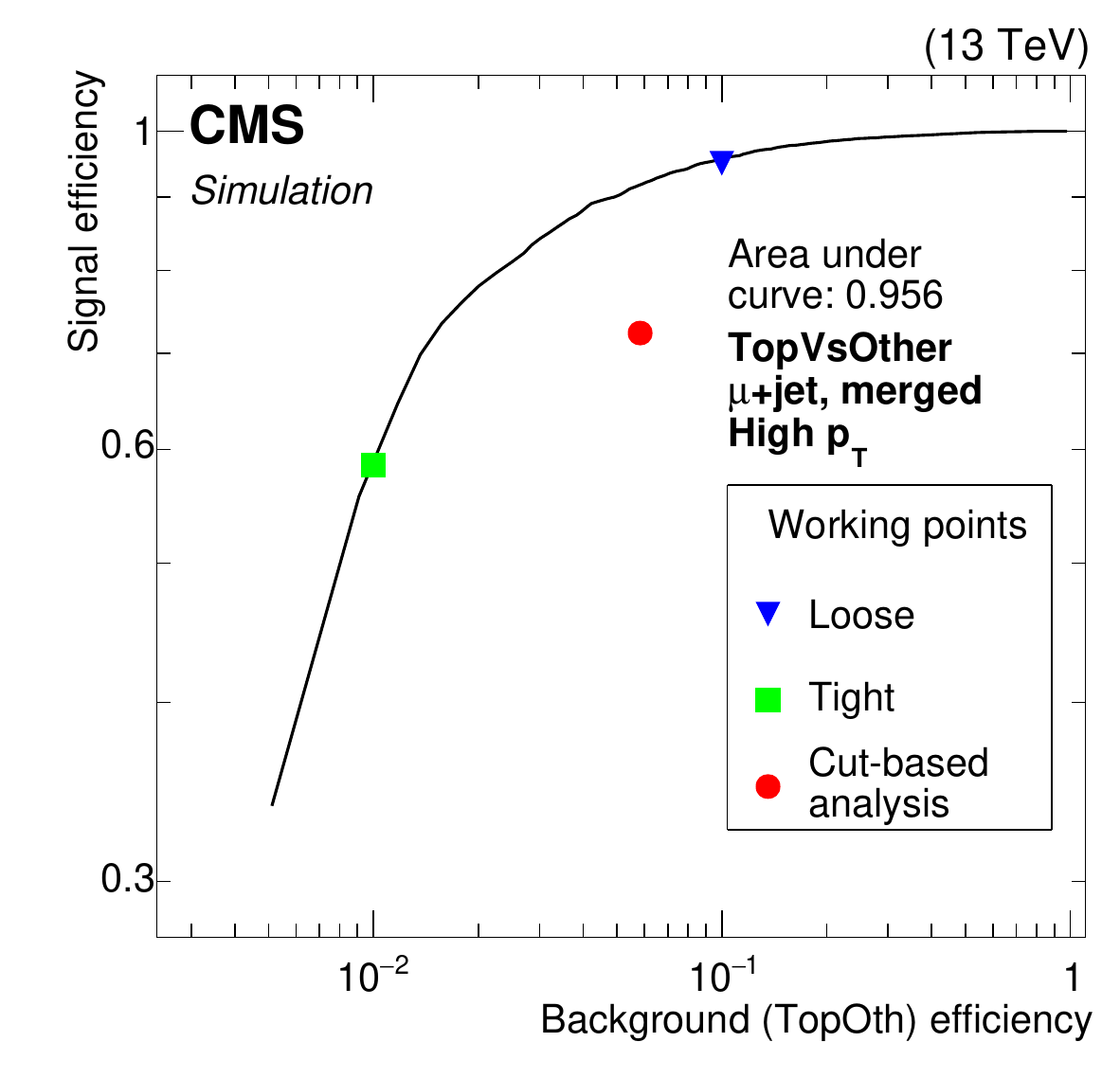}
\caption{The receiver operating characteristic curve for the TopVsQCD (left) and TopVsOther (right) top quark taggers. The signal efficiency and background misidentification rate are evaluated on top quark candidates with an associated muon, with $\pt > 500\GeV$ and in the merged configuration in simulated \ttbar events. The red triangle in the left plot corresponds to the requirement on the TopVsQCD variable providing a signal efficiency of 99\%; the blue triangle and the green square in the right plot correspond to the loose and tight WP of the TopVsOther tagger, respectively. A comparison with a cut-based selection (red point) is also shown in the TopVsOther case.}
\label{fig:topROC}
\end{figure}

The discrimination powers of the TopVsQCD and TopVsOther top quark tagging variables are illustrated in Fig.~\ref{fig:topROC}, showing their respective receiver operating characteristic curves.
A comparison with a cut-based selection is also shown; this selection requires the lepton to be isolated and applies a \PQb tagging score threshold with a background efficiency of 1\%. The top quark candidate for the cut-based approach is then chosen as the one with the lowest difference in mass with respect to the nominal top quark mass value taken from Ref.~\cite{PDG2024}. For a background misidentification rate of about 6\%, the selection based on the BDT gives a relative improvement on the true top quark candidate identification efficiency of about 30\%.

Top quark candidates are required to be above a threshold of the TopVsQCD score,
with a signal efficiency of 99\%, and to satisfy the loose or the tight WP requirement
of the TopVsOther variable, which respectively correspond to
a background rejection rate of 90 and 99\%. The signal efficiency and background misidentification
rate are evaluated on top quark candidates with an associated muon, with $\pt > 500\GeV$ and in the
merged configuration in simulated \ttbar events.

Events are selected if at least one top quark candidate
passes the loose WP of the TopVsOther tagger.
Events with no loosely tagged top quark candidates are still selected
if the lepton satisfies tight
identification requirements (referred to in the following as ``tight lepton''); these events are
employed to calibrate the $\Xbb$ tagging as described in Section~\ref{sec:background}.

\section{Event categorization}\label{sec:categories}
Selected events are assigned to mutually exclusive categories,
which are used for signal extraction
or as background control regions (CRs).
Where there are multiple reconstructed top quark
or $\PGf$ boson candidates in an event, a procedure is defined
to select only one of them, as described below.

\begin{figure}[ht!]
\centering
\includegraphics[width=0.59\linewidth]{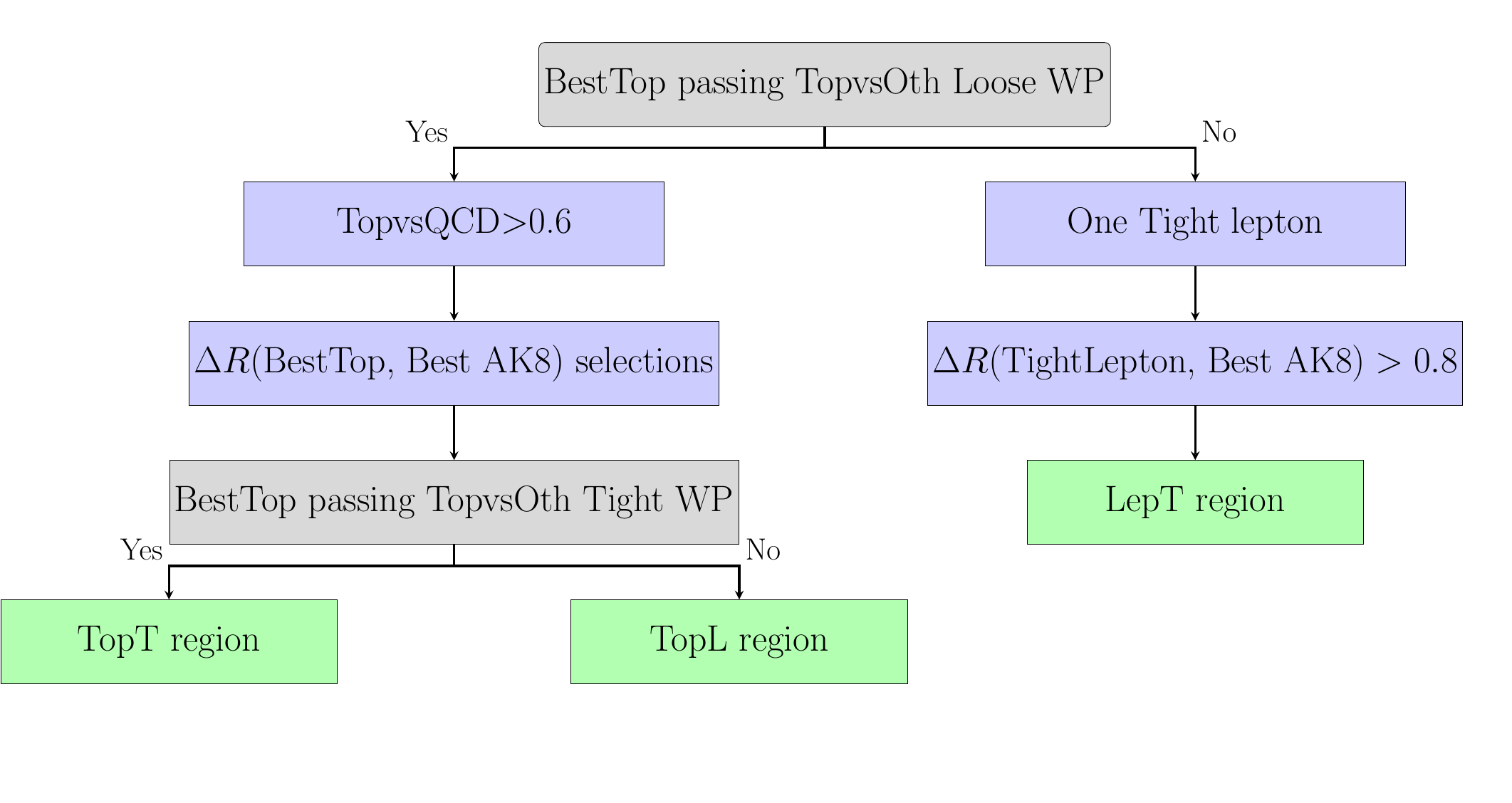}
\hfill
\includegraphics[width=0.40\linewidth]{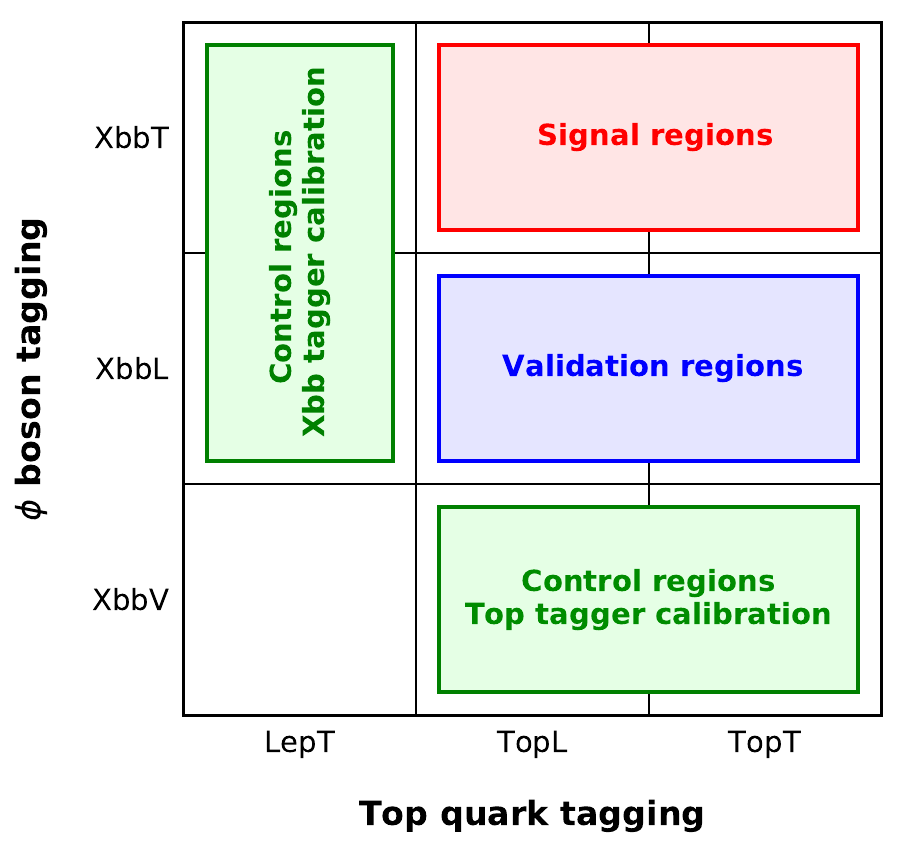}
\caption{Left: block diagram showing the event classification based on top quark tagging. Right: schematic view of the signal, validation, and control regions defined in the analysis. The label BestTop (Best AK8) refers to the top quark candidate (AK8 jet) with the highest value of the TopVsOther (XbbVsQCD) tagger. The labels TopT (Top Tight) and TopL (Top Loose) refer to events where there is a top quark passing a tight working point for the TopVsOther tagger with a background rejection of 99\%, or a loose working point with a background rejection of 90\% while vetoing the tight working point to ensure statistical independence of the categories. The label LepT refers to events that are neither in the TopT or TopL categories, but where a top quark candidate is built via a lepton and b-jet pair. The XbbT, XbbL, and XbbV labels refer to the fact that the largest score of the XbbVsQCD tagger evaluated on AK8 jets selected in the event is in the XbbVsQCD range of (${>} 0.98$), (0.80--0.98), or (${<} 0.80$), respectively. Additionally, for the event to be included in the XbbV category, the mass of the jet with the largest score is required to be in the 60--220\GeV range.}
\label{fig:regions}
\end{figure}

Three exclusive event categories are defined, based
on the tagging quality of the reconstructed top quark candidates.
A block diagram illustrating the workflow to assign an event
to one of these three top quark tagging categories is shown in
Fig.~\ref{fig:regions}~(left).
An event is assigned to the TopT category
if there is at least one top quark
candidate passing the tight WP requirement of the TopVsOther tagger.
Events with no top quark candidates passing the
above selection, but with at least one candidate passing
the loose WP of the TopVsOther tagger, 
are assigned to the TopL category.

If there are multiple top quark candidates passing the tight (loose) WP
in the TopT (TopL) category,
they are ordered as follows, and the highest in the order is selected
as the top quark candidate in the event.
The highest priority is given to a high-\pt top quark candidate
in the merged configuration,
then to a low-\pt (then high-\pt) candidate in the resolved configuration,
and lastly to a low-\pt candidate in the merged configuration.
Top quark candidates with the same priority are ordered by
descending values of the TopVsOther tagger output.

Events not assigned to any of the above categories
are required to contain one tight lepton
and assigned to the LeptonTight (LepT) category.

In order to further reduce the \ttbar and QCD multijet background,
the events must also pass the following requirements:
\begin{itemize}
\item {\tolerance=800 for events in the TopT or TopL category, the top quark candidate must have $\mathrm{TopVsQCD} > 0.6$;\par}
\item to exploit the fact that the top quark and $\PGf$ boson are produced mainly back-to-back in the CMS rest frame,
there must be at least one AK8 jet that is separated from the top quark candidate by $\Delta R >1.2$,
from the AK4 jet of the top quark candidate by $\Delta R >1.2$,
and from the lepton of the top quark candidate by $\Delta R >0.8$;
\item for events in the LepT category, there must be at least one AK8 jet that is
separated from the tight lepton of the event by $\Delta R >0.8$.
\end{itemize}

The AK8 jet with the largest value of the XbbVsQCD tagger, selected among the ones passing the above $\Delta R$ requirements,
is chosen as the $\PGf$ boson candidate of the event and, depending on the value of the tagger output,
the event is assigned to one of three exclusive categories as follows:
if $\mathrm{XbbVsQCD}>0.98$ or $0.80<\mathrm{XbbVsQCD}<0.98$, the event is assigned
to the XbbTight (XbbT) or XbbLoose (XbbL) category, respectively;
if $\mathrm{XbbVsQCD}<0.80$ and the AK8 jet mass
is in the 60--220\GeV range, the event is assigned to the
XbbVeto (XbbV) category; otherwise the event is discarded.

Nine independent event regions are thus defined,
based on all the possible combinations of the three
top quark tagging categories (TopT, TopL, LepT) and the three
$\Xbb$ tagging categories (XbbT, XbbL, XbbV) the event may belong to, as shown in Fig.~\ref{fig:regions}~(right).

{\tolerance=800 Events in the (TopT, XbbT) and (TopL, XbbT) regions, which
have the largest signal-to-background ratio across all regions,
are used for signal extraction (signal regions, SRs).
Events in the signal-depleted regions
(TopT, XbbL) and  (TopL, XbbL) are employed
as validation regions (VRs) to verify that good agreement is
found between data and background prediction,
and to constrain the background in the final fit.
The events in each of these regions are further divided in
subcategories, depending on the flavor of the lepton
(muon or electron) of the top quark candidate and the
presence or absence of at least one forward jet, to exploit the presence in signal events of high-pt quarks produced at low angle with respect to the beam axis.\par}

For each event belonging to the SRs or VRs,
a \PQT quark candidate is reconstructed by summing together the four-momenta of the reconstructed 
top quark candidate and of the AK8 jet chosen as the $\PGf$ boson candidate.
The distribution of the invariant mass of the \PQT quark candidate $\MlvbJ$ is then used in the final fit for signal extraction.

The regions (TopT, XbbV) and (TopL, XbbV),
as well as the regions (LepT, XbbT) and (LepT, XbbL),
are employed as CRs for top quark tagging
and $\Xbb$ tagging calibration, respectively.
The signal-to-background ratio in the CRs is lower by a factor of 15--30 compared to the SRs.

\section{Background estimation}\label{sec:background}
After the event selection, the major sources of background in the SRs are mainly \ttbar events, followed by single top quark events, and, to a lesser extent, \ttH, QCD multijet and \wjets events. Other sources, such as diboson processes and top quarks produced in association with vector bosons, are found to be negligible both in the SRs and CRs. 

The event rates and distribution shapes used as input to the final fit
for signal extraction are derived from
simulated events. Data-to-simulation correction factors
are applied to simulated events in order to correct small discrepancies
between data and simulation
in the trigger and reconstruction efficiencies,
and for the lack of higher-orders corrections in the simulation.

A relevant aspect of the analysis is the usage of
top quark and $\Xbb$ tagging to identify the top quark and
$\PGf$ boson candidates, respectively. Data-to-simulation correction
factors for the tagger efficiency and mistag rate are determined
from the CRs, and applied to simulated signal and background events.
The corrections correspond to variations from the predicted efficiency or mistag rate
in the range between 5 and 15\%.

Events in the (TopT, XbbV) and (TopL, XbbV) regions have a top quark candidate
satisfying top quark tagging criteria. They are employed to calibrate the
top quark tagger efficiency and mistag rate by means of a background-only fit to the \PQb~$\mathrm{tag}(j)$
distribution of the AK4 jet associated to the top quark candidate.
In the fit, the templates of the background distributions
are divided in three components, 
according to the type of the reconstructed top quark candidate
(\Ptrue, \PQCD, or
\POth), as inferred from generator-level information,
and the top quark tagging correction factors are determined from the fitted
yields of these components.
The fit takes into account the correlation of
event yields in the TopL and TopT categories,
since a change in the top quark tagging efficiency may
cause migration of events from one category to the other. 
Different correction factors are determined for events in the muon and electron channels.
All the systematic effects, which are described
in Section~\ref{sec:systematics}, are included
in the fit by means of nuisance parameters.

\begin{figure}[htbp!]
    \centering
    \includegraphics[width=0.49\textwidth]{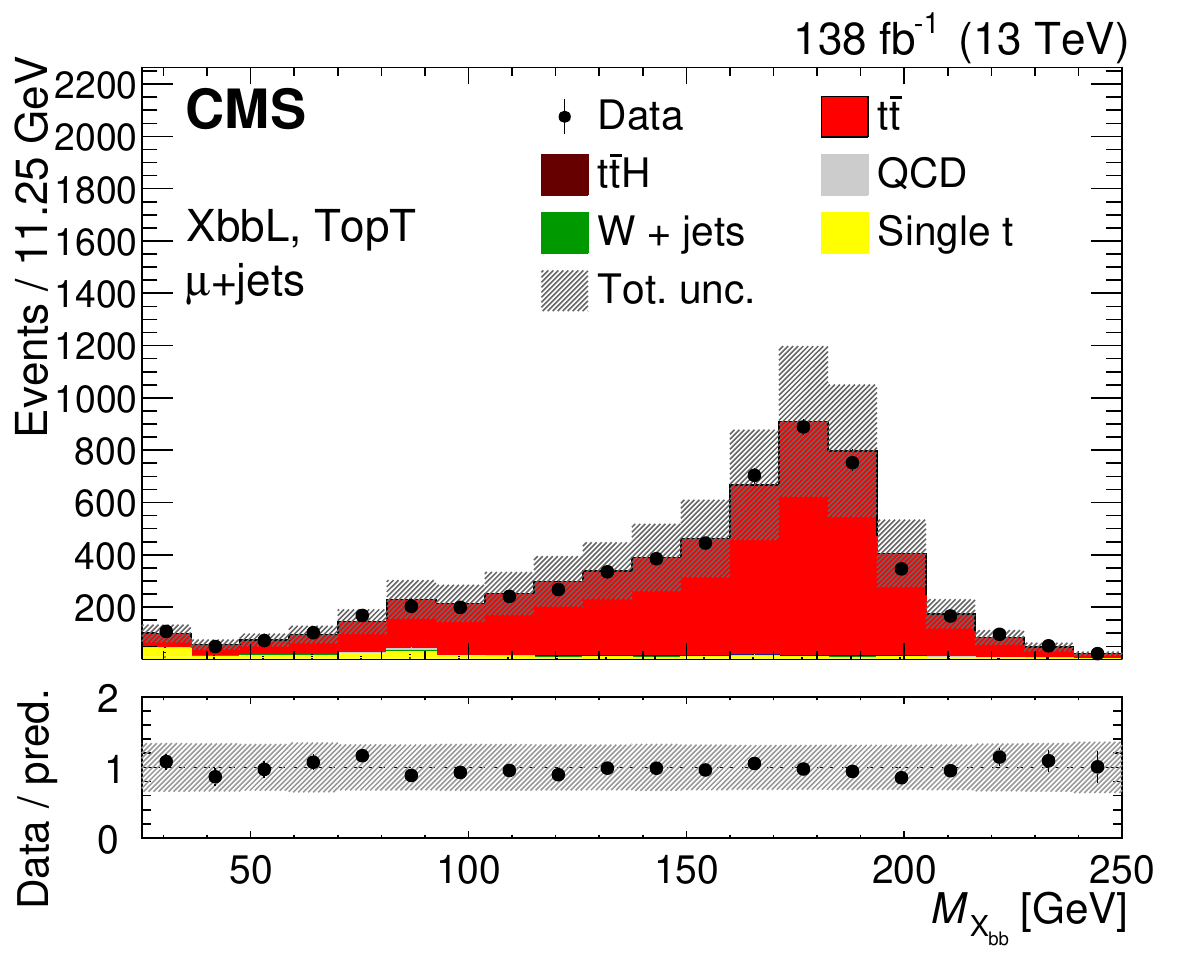}
    \includegraphics[width=0.49\textwidth]{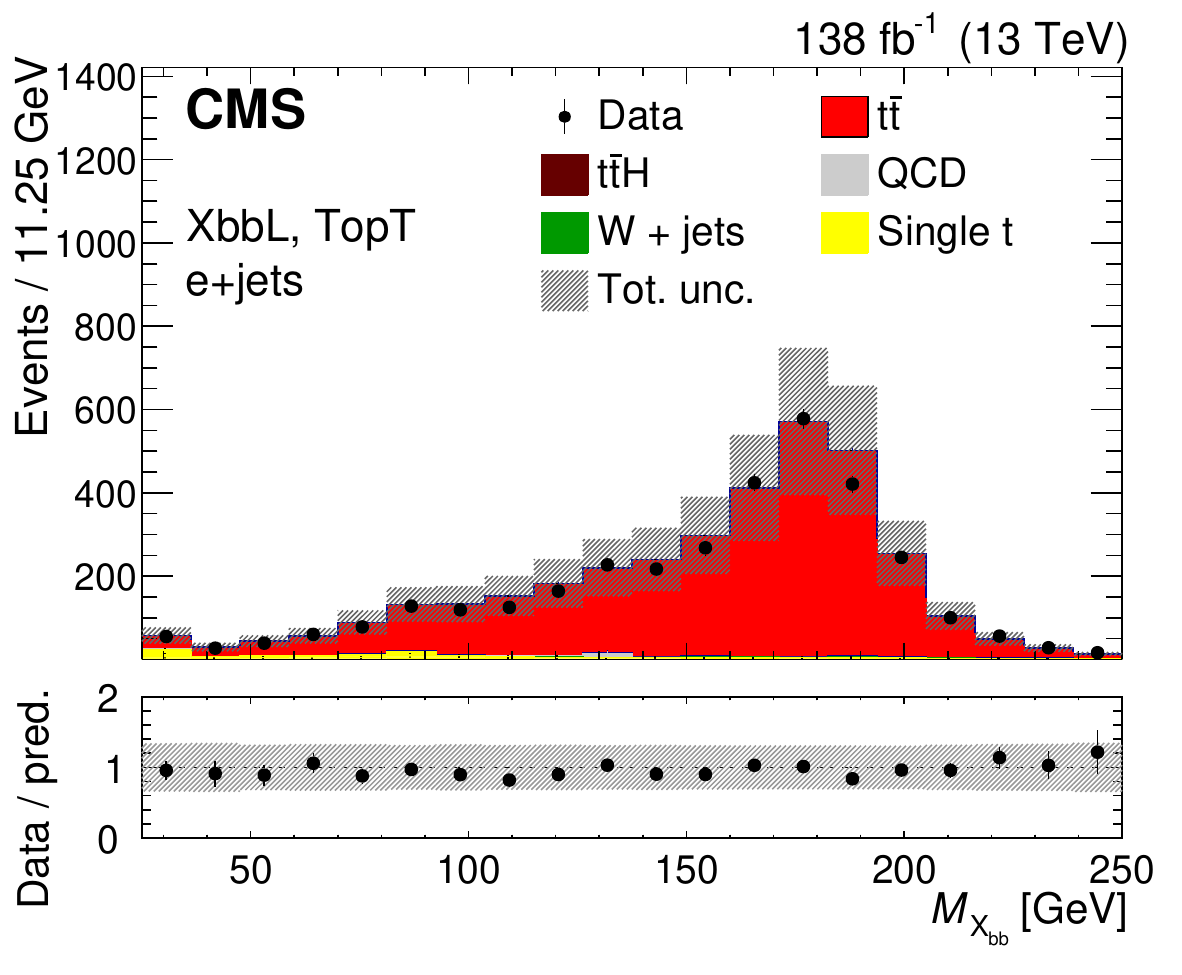}
    \includegraphics[width=0.49\textwidth]{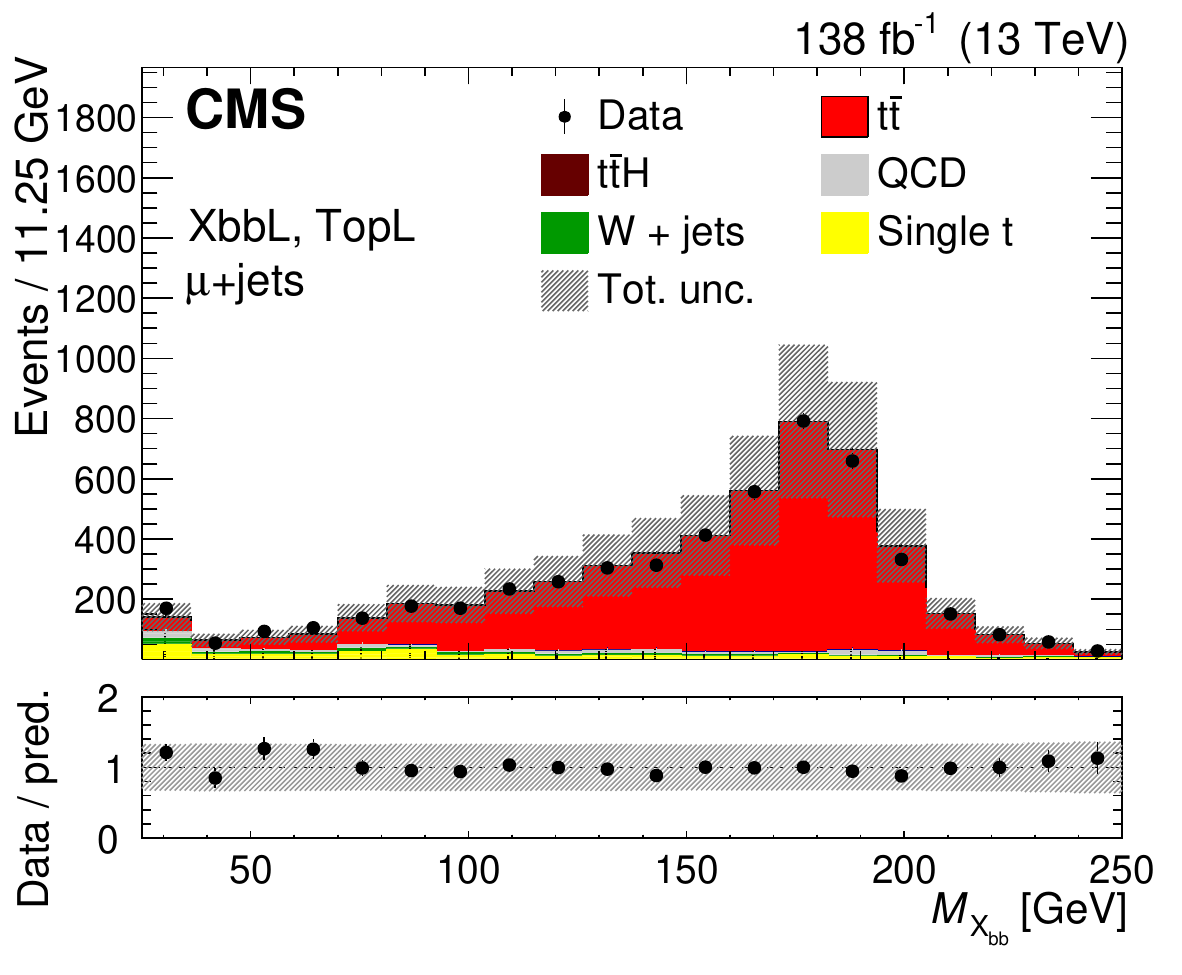}
    \includegraphics[width=0.49\textwidth]{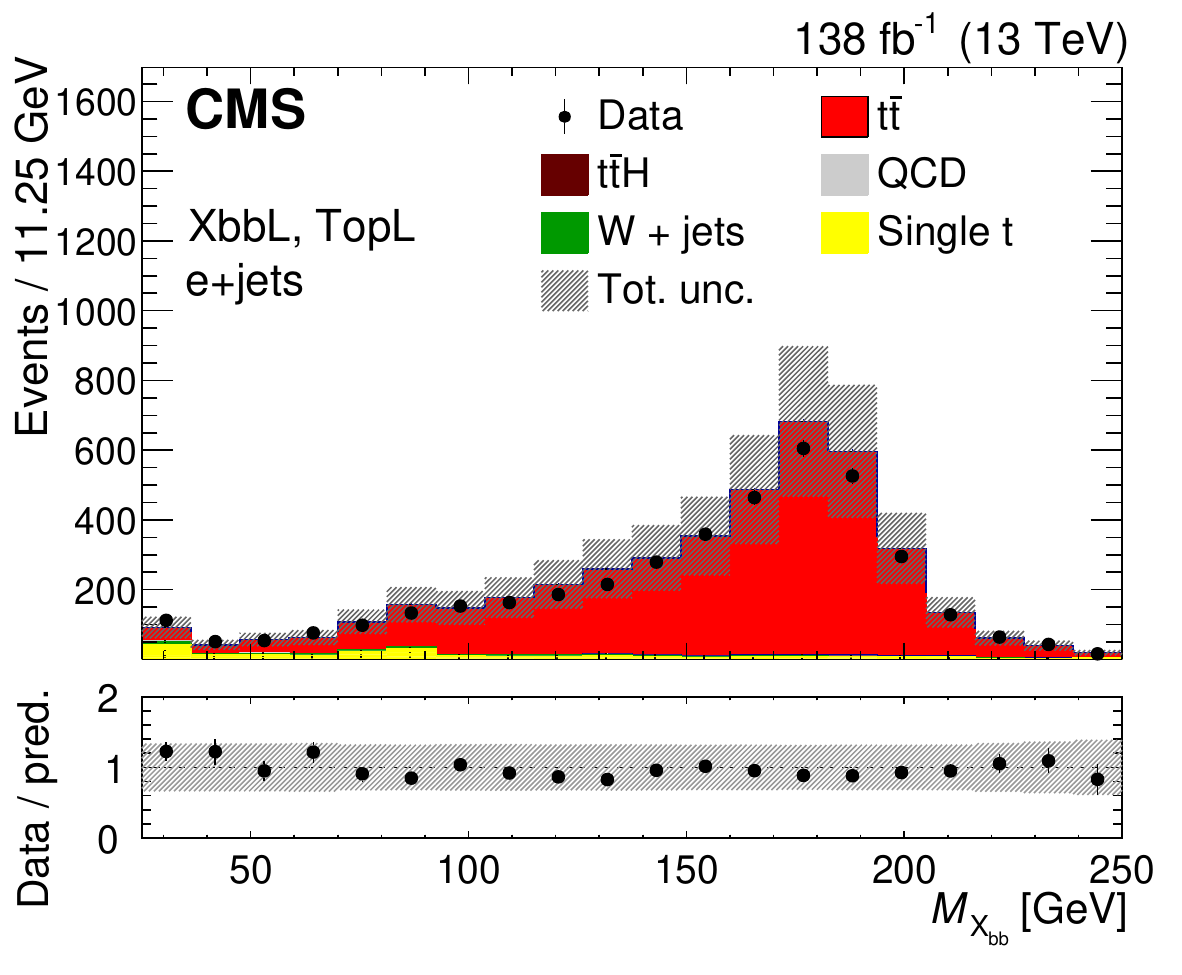}
    \caption{Distribution of the \MXbb mass of the AK8 jet selected as the $\PGf$ boson
      candidate for data and simulated background events
      in the (TopT, XbbL) (upper), in the (TopL, XbbL) (lower) VRs,
      for the muon (left) and electron channels (right).
      The distributions are shown before the final fit for signal extraction.}
    \label{fig:validregions_Run2}
\end{figure}

Background events in the (LepT, XbbT) and (LepT, XbbL) regions
have an AK8 jet misidentified as a $\PGf$ boson candidate and
are employed to calibrate the mistag rate
of the $\Xbb$ tagger.
Correction factors are determined from the fitted
event yields of the background components
in a fit to the distribution of reconstructed mass \MXbb of the
$\Xbb$ jet selected as the $\PGf$ boson candidate. 
Only events with $\MXbb > 110\GeV$ are considered for
the fit, since in this mass range the background compositions
of the CRs agrees with those of the SRs and VRs.
The templates of the background distributions are
divided in two components: resonant background events from all-hadronic top quark decays
and nonresonant background events.
The fit	takes into account the correlation of
event yields in the XbbL and XbbT categories,
since a change in the $\Xbb$ tagging efficiency may
cause migration of events from one category to the other.
All the systematic effects, which are described
in Section~\ref{sec:systematics}, are included
in the fit by means of nuisance parameters.
As a proof that the mistag correction factors are independent of
\MXbb, Fig.~\ref{fig:validregions_Run2} shows
the \MXbb distribution for events in the VRs (TopT, XbbL) and (TopL, XbbL), with
all the correction factors applied and before the final fit for signal extraction.

The efficiency of the $\Xbb$ tagger to identify correctly an $\Xbb$
jet is calibrated using data with energetic gluons splitting
to bottom quarks~\cite{CMS-PAS-BTV-22-001}. Data-to-simulation correction
factors are derived in bins of the AK8 jet \pt and applied to simulated
signal events and \ttH background events.

\clearpage

\section{Systematic effects and uncertainties}\label{sec:systematics}
Systematic uncertainties, both of theoretical and experimental origin,
affect the event rate of the background and signal predictions,
the shape of $\MlvbJ$ invariant mass distribution, or both.
Depending on the source, the uncertainties are treated as 
fully correlated or fully uncorrelated among the different
data-taking periods, or decomposed into a correlated and
an uncorrelated component.

Most of the uncertainties are related to the corrections that are applied to simulated
events in order to account for the known discrepancies with data.
In this case, and for each systematic correction,
modified signal and background distributions 
are obtained by shifting the nominal value of
the correction up and down by one standard deviation.
A summary of the systematic uncertainties
is reported in Table~\ref{tab:systematics}.

The integrated luminosity of the entire 2016--2018 dataset has an uncertainty
of 1.6\%~\cite{CMS-LUM-17-003,CMS-PAS-LUM-17-004,CMS-PAS-LUM-18-002},
which applies to the overall event rates of all simulated processes. 
 
The uncertainty associated with the mismodeling of pileup is evaluated by varying the total inelastic $\Pp\Pp$ cross section used
to compute the pileup distribution in data by $\pm$4.6\%~\cite{Sirunyan:2018nqx}. 

Lepton and jet trigger efficiencies are measured in data and simulated samples as a function of lepton \pt and $\eta$, and AK8 jet \pt and mass.
Data-to-simulation scale factors (SF) are then obtained and applied to simulated events.

During the 2016--2017 data taking, a gradual shift in the timing of the inputs of the ECAL L1 trigger in the region $\abs{\eta} > 2.0$ caused a specific trigger inefficiency. For events containing a jet
with $\pt \gtrsim 100\GeV$ in the region $2.5 < \abs{\eta} < 3.0$, the efficiency loss is
approximately 10--20\%, depending on \pt, $\eta$, and data-taking period~\cite{CMS:2020cmk}.
Correction factors are computed from data and applied to the acceptance evaluated from simulation.

Differences between data and simulation in the efficiency of
lepton reconstruction and identification (ID)
are corrected by applying SFs, which are functions of lepton \pt and $\eta$, to simulated events.

The SFs are also applied to simulated events in order
to reproduce the efficiency of the pileup BDT discriminator cut applied to AK4 jets
with $30<\pt<50\GeV$ to remove jets originating from pileup.

Systematic effects due to differences in the jet energy scale between data and simulation
are evaluated by varying the jet four-momenta in simulated events, according
to the \pt and $\eta$-dependent uncertainties in the jet energy scale. 
Similarly, in order to evaluate systematic effects due to differences in the jet energy resolution
between data and simulation, a smearing is applied to simulated events
by increasing or decreasing the jet resolutions by their uncertainties~\cite{CMS:2016lmd}.
The variations in jet four-momenta are also propagated to the \ptvecmiss.

Correction factors for the top quark and $\Xbb$ tagging
are derived from fits to data in the CRs, as described in Section~\ref{sec:background},
and by using data events with energetic gluons splitting to bottom quarks~\cite{CMS-PAS-BTV-22-001}.
The uncertainties in the correction factors are taken as the systematic uncertainties
in the top quark and $\Xbb$ tagging.

The uncertainties from the choice of the PDFs are evaluated by reweighting
the simulated signal and background events using the NNPDF~3.1
Hessian uncertainties~\cite{Butterworth:2015oua}.

Simulated \ttbar events are corrected by applying
event weights derived from the ratio of top quark \pt, measured in data
at parton level, and NNLO predictions~\cite{Khachatryan:2016mnb}.
An uncertainty of ${\pm} 50 \%$ of the applied correction is assigned to this
event reweighting.

The choice of the factorization and renormalization scales, $\mu_\mathrm{F}$ and $\mu_\mathrm{R}$, affects both the cross section and the acceptance of each background. The uncertainties are estimated separately for each background by halving or doubling each of the scales, compared to the nominal values used in simulation.

The statistical uncertainty due to the limited size of simulated event samples is evaluated for each bin with
the Barlow--Beeston ``lite'' method~\cite{Barlow:1993dm, Conway:2011in}. The impact of this systematic effect on the result of the fit is large in the high-mass bins of the $\MlvbJ$ distribution, in particular above 1.5\TeV, due to the fall-off of standard model background processes. This systematic uncertainty is also not reduced by the simultaneous fit to VR and SR, since it is uncorrelated across bins and regions.

\begin{table}[!ht]
  \topcaption{List of systematic uncertainties, showing whether a source modifies the event rate or the distribution shape, and whether the effect is
    correlated or uncorrelated across the years of data taking. All the listed sources affect both signal and background processes. The relative impact on the postfit signal strength is shown for a representative set of signals with a \PQT mass  $(m_{\PQT})$ hypothesis of 1.8\TeV and varying the mass of $\PGf$ boson.}
 \label{tab:systematics}
 \centering
 \resizebox{\textwidth}{!}{
 \begin{tabular}{l c c c}
 Source & Rate/shape & Correlation & Impact ($m_{\PQT} = 1.8\TeV$) [\%] \\
 \hline
 $\Xbb$ tagging & shape & correlated + uncorrelated & 20--90 \\
 Top quark tagging & shape & correlated + uncorrelated & 10--40 \\
 Simulated samples size & shape & uncorrelated & 40--60 \\
 $\mu_\mathrm{R}$, $\mu_\mathrm{F}$ & shape & correlated & 10--20 \\
 Parton distribution functions & shape & correlated & 5--8\\
 Top quark \pt reweighting & shape & correlated & ${<} 1$\\
 Pileup jet identification & shape & correlated & 5--15\\
 ECAL L1 trigger inefficiency & shape & correlated & 4--5 \\
 Jet energy scale & shape & uncorrelated & 3--5\\
 Jet energy resolution & shape & uncorrelated & 3--5\\
 Jet pileup identification & shape & correlated & ${<} 1$\\
 Trigger & shape & uncorrelated & 2--5 \\
 Lepton reconstruction and ID & shape & correlated + uncorrelated & 2--5\\
 Integrated luminosity & rate & correlated & 1--2 \\
 \end{tabular}
 }
\end{table}

\section{Results}\label{sec:results}
The signal extraction procedure is based on a simultaneous
maximum likelihood fit to the invariant mass distribution
of the \PQT quark candidates $\MlvbJ$
in the (TopT, XbbT) and (TopL, XbbT) SRs, and
(TopT, XbbL) and (TopL, XbbL) VRs.
The events are divided into 16 categories defined according to the
specific SR or VR, the lepton channel, and the presence
or absence of at least one forward jet.
Systematic uncertainties are treated in the fit as nuisance parameters, assumed to be described by a log-normal
probability density function for those that affect the rate, or a Gaussian
probability density function for those that also affect the shape of the distributions.

\begin{figure}[ht!]
\centering
\includegraphics[width=.48\textwidth]{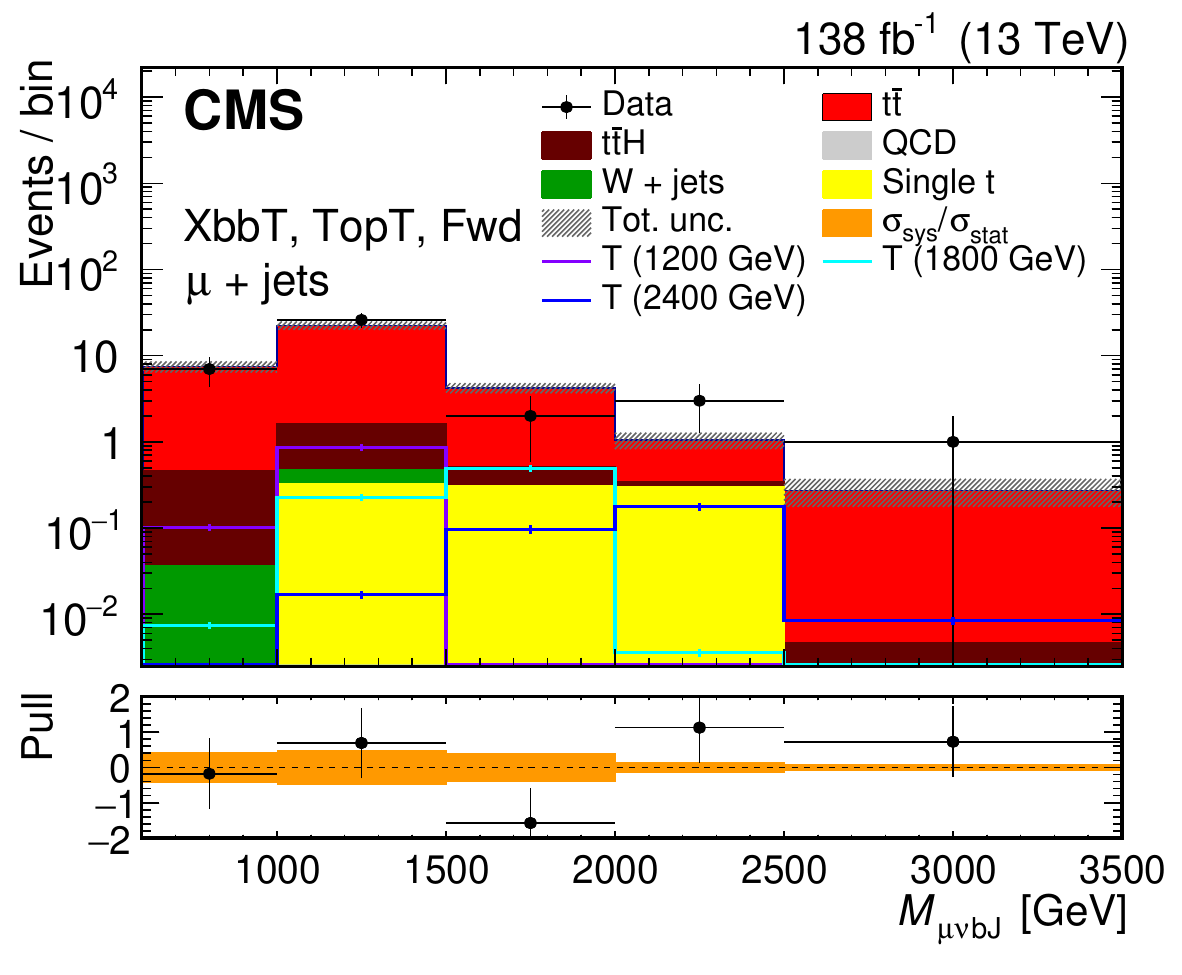}
\includegraphics[width=.48\textwidth]{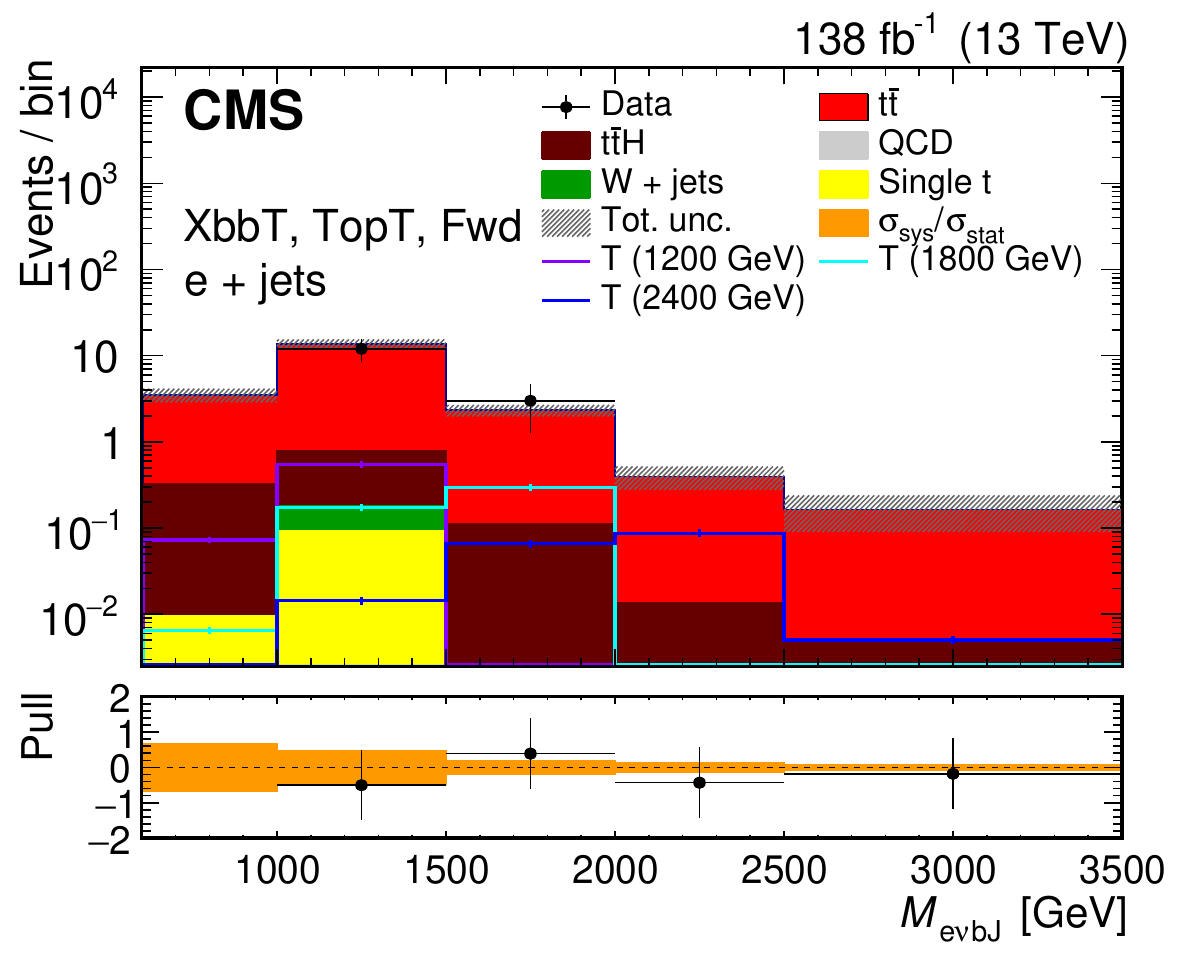}
\includegraphics[width=.48\textwidth]{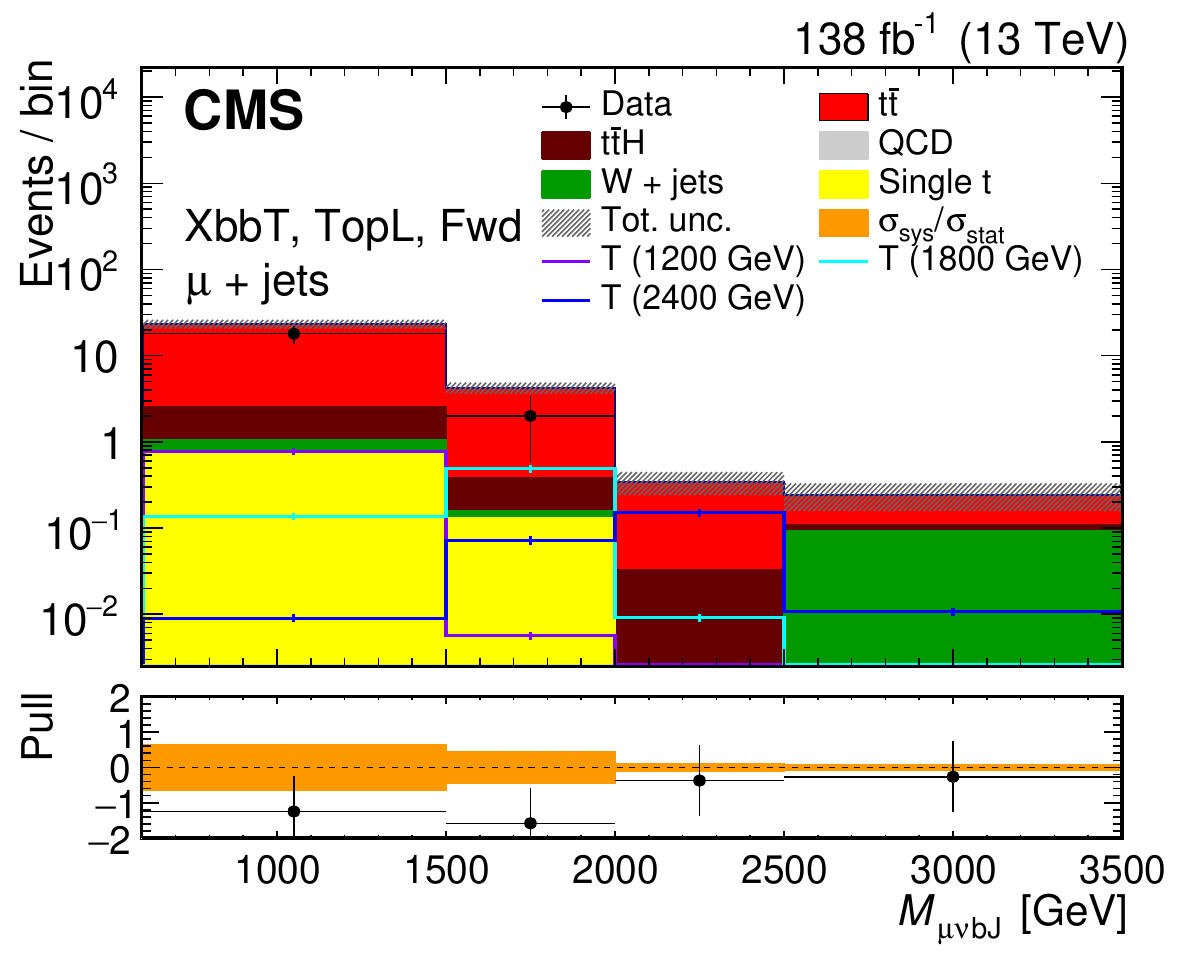}
\includegraphics[width=.48\textwidth]{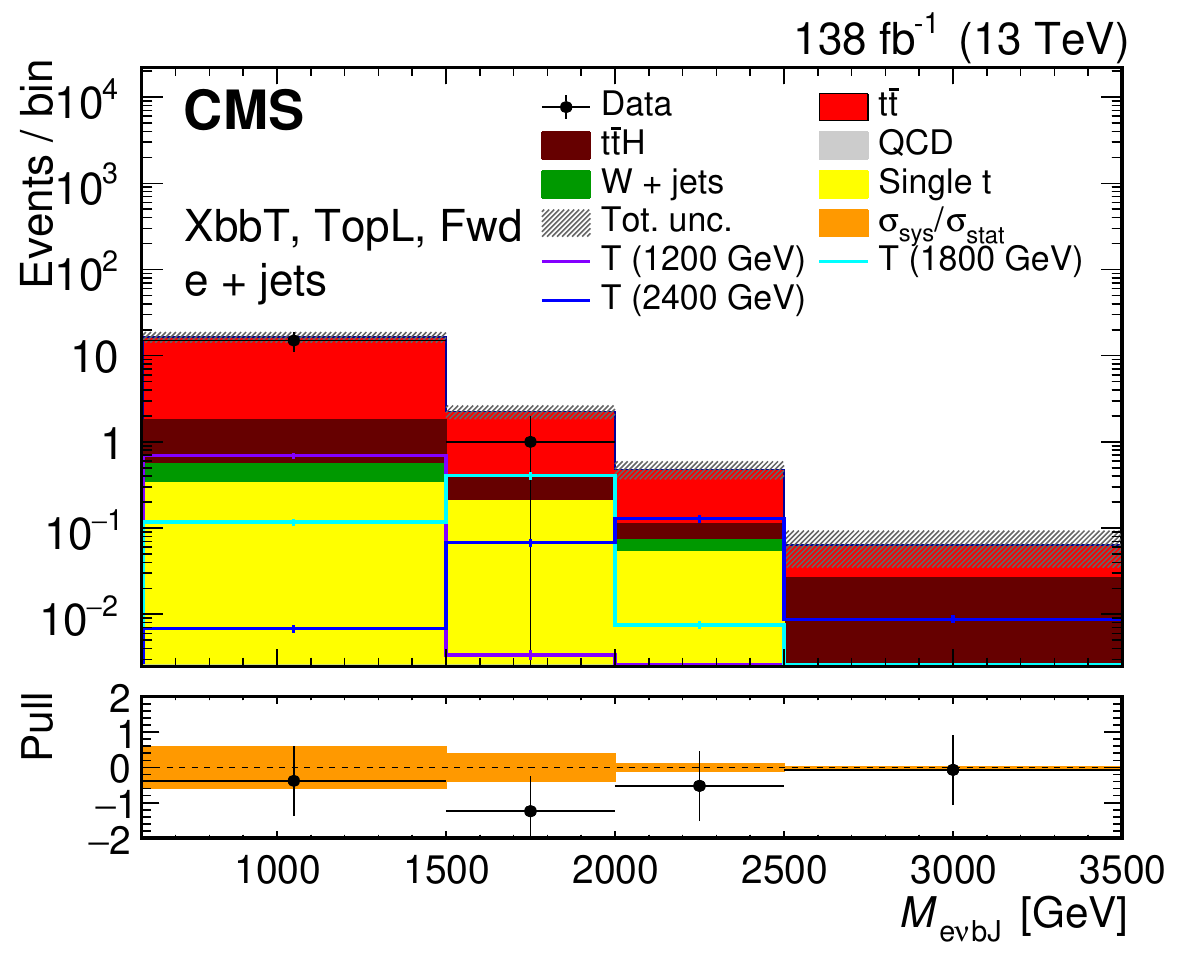}
\caption{Invariant mass distribution of the $\PQT$ quark candidates selected in the $\PQT\to\PQt\PH$ channel, for events with at least one forward jet and in the SRs for (TopT, XbbT) (upper) and (TopL, XbbT) (lower). For these events, the reconstructed mass of the Higgs boson candidate is required to be between 110 and 140\GeV. Distributions are shown for events in the muon (left) and electron (right) channel. The first (last) bin of each distribution also includes underflow (overflow) events. Background distributions are normalized to the results of the background-only fit, while signal distributions are overlaid, normalized to the expected cross section computed at NLO for a singlet \PQT quark with a resonance width $\Gamma$ of 1\% of the mass. The lower panels show the difference between the number of events in data and that expected from the background-only fit, normalized to the statistical uncertainty of the data. When no events are observed in data, the Poissonian upper limit at 68\% \CL is used for the calculation of the pull. The orange band represents the systematic uncertainties, also normalized to the statistical uncertainty of the data.}
\label{fig:sigregiondistr}
\end{figure}

Events used as input to the fit must satisfy a requirement on the
reconstructed $M_{\Xbb}$ mass of the $\PGf$ boson candidate,
which is defined such that it selects 95\% of signal events
and depends on the $\PGf$ boson mass ($m_{\PGf}$) hypothesis
being tested, as shown in Table~\ref{tab:phimasscut}.
The requirement is asymmetric with respect to the $m_{\PGf}$ value,
because of the asymmetric tails of the $M_{\Xbb}$ mass distribution,
with the left tail becoming broader as $m_{\PGf}$ increases.
\begin{table}[!ht]
\topcaption{Requirement on the reconstructed $M_{\Xbb}$ mass of the
   $\PGf$ boson candidate, as a function of the $m_{\PGf}$ boson mass hypothesis
  being tested. The $M_{\Xbb}$ mass must fall within the specified interval.}
\label{tab:phimasscut}
\centering
\renewcommand{\arraystretch}{1.1}
\begin{tabular}{l c}
 $m_{\PGf}$ [{\GeVns}] & $M_{\Xbb}$ requirement [{\GeVns}] \\ [\cmsTabSkip]
 \hline
$25$  & $[20, 34]$   \\
$50$  & $[41, 73]$   \\
$75$  & $[60, 100]$  \\
$100$ & $[78, 130]$  \\
$125$ & $[97, 160]$  \\
$150$ & $[120, 190]$ \\
$175$ & $[140, 210]$ \\
$200$ & $[160, 230]$ \\
$250$ & ${>} 190$ \\
 \end{tabular}
\end{table}
In the hypothesis of the SM Higgs boson ($\PQT\to\PQt\PH$ decay),
the requirement $110 < M_{\Xbb} < 140\GeV$ is imposed,
to optimize the analysis sensitivity.
Figure~\ref{fig:sigregiondistr} shows the fitted $\MlvbJ$
distributions of the \PQT quark candidates for events in the SRs (TopT, XbbT), and (TopL, XbbT), with at least one forward jet,
and passing the $M_{\Xbb}$ selection of the $\PQT\to\PQt\PH$ decay analysis.

The following results have been determined using \textsc{Combine}~\cite{CMS:2024onh},
the CMS statistical analysis tool,
which is based on the \textsc{RooFit}~\cite{Verkerke:2003ir}
and \textsc{RooStats}~\cite{Moneta:2010pm} frameworks.

\begin{figure}[ht!]
\centering
\includegraphics[width=1.0\textwidth]{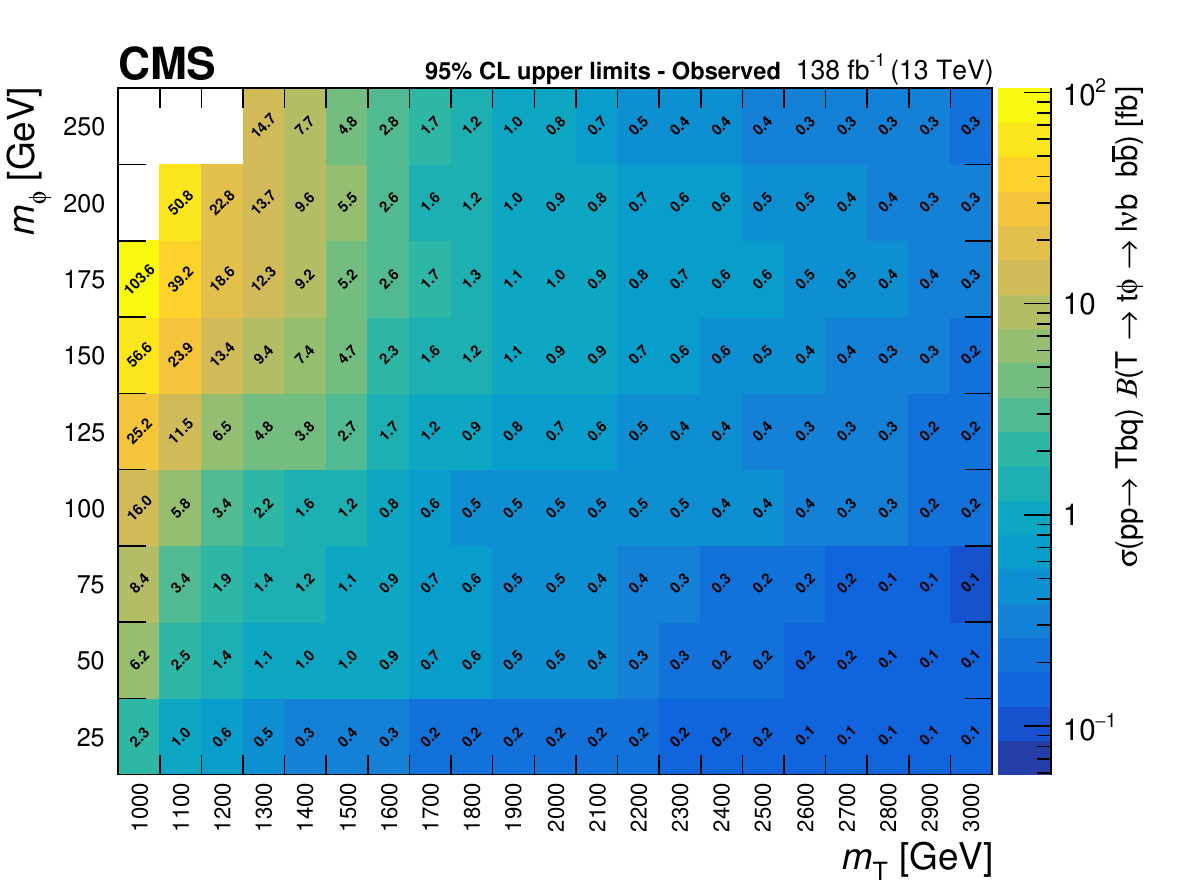}
\caption{Observed 95\% \CL upper limits on the single \PQT quark product of cross section and
  branching ratio for the $\PQT\to\PQt\PGf \to \PQb\Pell\PGn \bbbar$ channel, as a function of
  $m_{\PQT}$ and $m_{\PGf}$ masses.}
\label{fig:2dlim}
\end{figure}

No evidence of a statistically significant signal is found from the fit to the data.
An upper limit at 95\% confidence level (\CL) is set on the product of the production
cross section and $\PQT \to \PQt\PGf \to \PQb\Pell\PGn\bbbar$ branching fraction
as a function of the \PQT quark mass ($m_{\PQT}$) and $m_{\PGf}$ hypotheses, as shown in Fig.~\ref{fig:2dlim}.
The limits are determined using the \CLs method~\cite{Junk:1999kv,Read:2002hq,Cowan:2010js,ATLAS:2011tau},
taking the profile likelihood ratio as the test statistic with the asymptotic limit approximation.
For a fixed $m_{\PQT}$, the upper limit tends to increase with $m_{\PGf}$, because the looser requirement
applied on $M_{\Xbb}$ increases the amount of the selected background.

\begin{figure}[ht!]
    \includegraphics[width=0.48\textwidth]{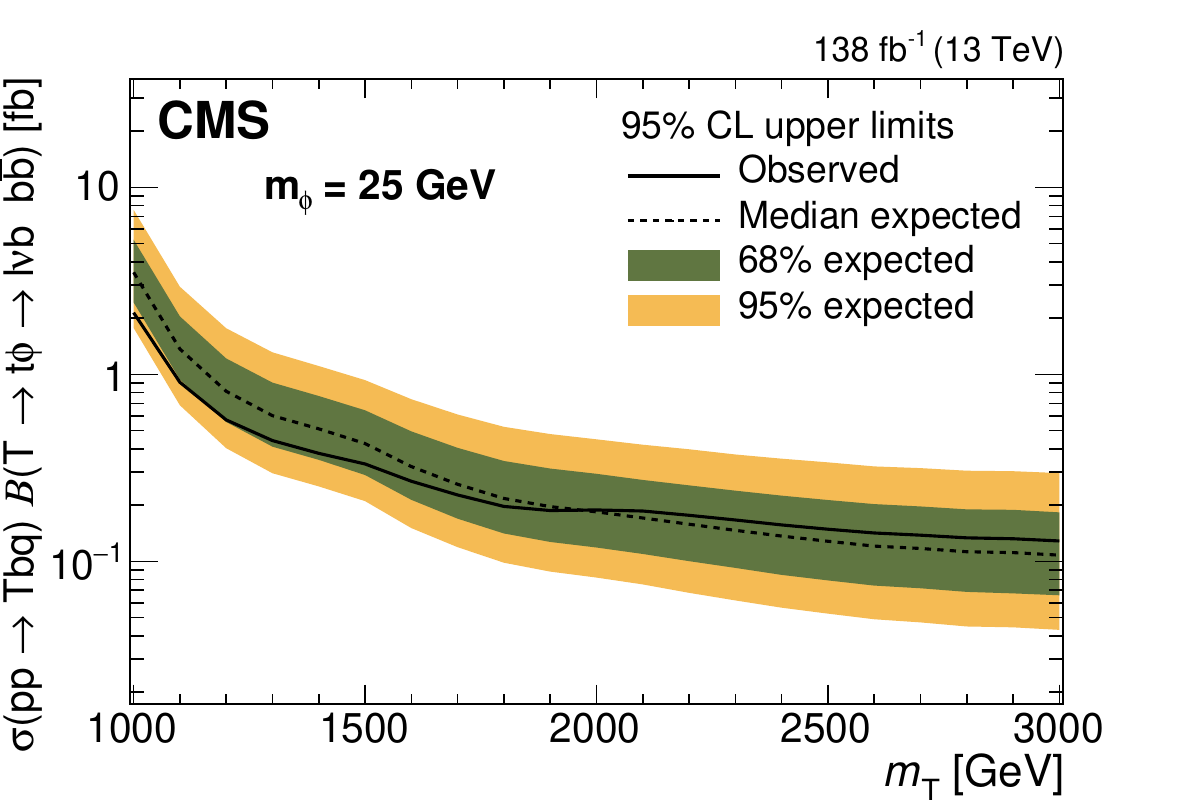}
    \includegraphics[width=0.48\textwidth]{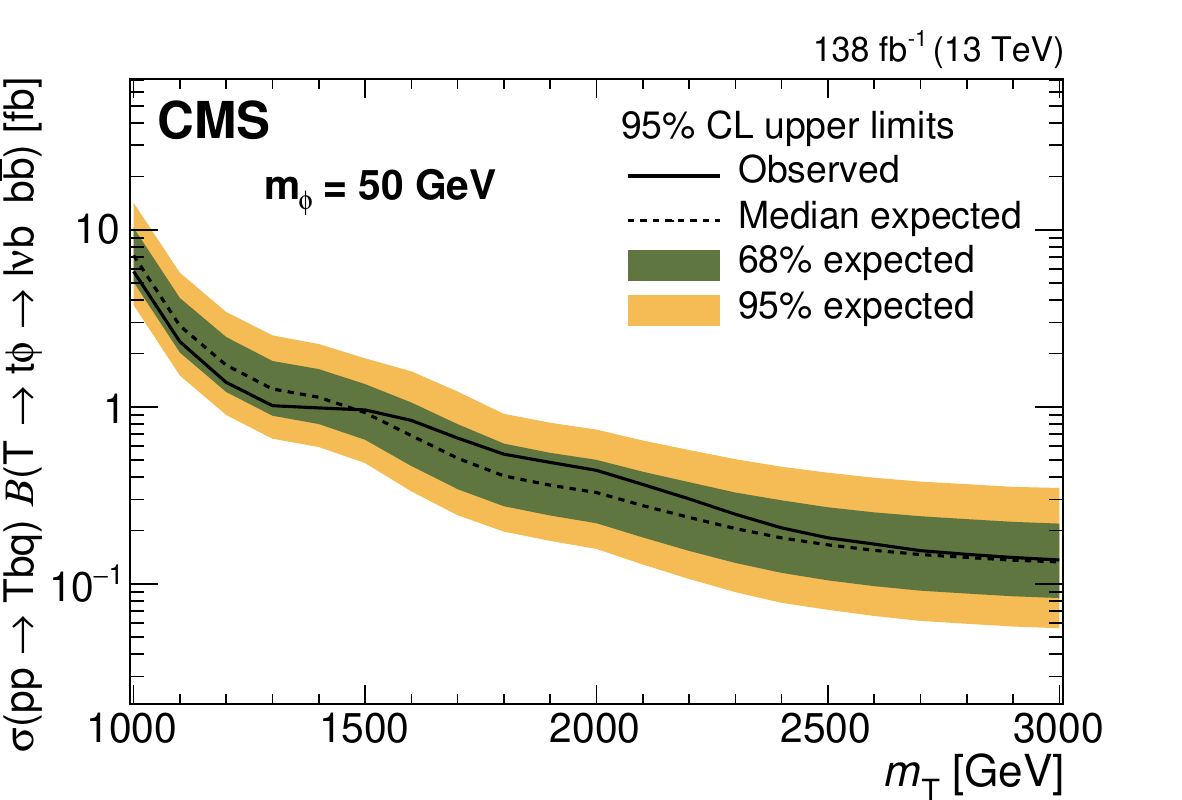}\\
    \includegraphics[width=0.48\textwidth]{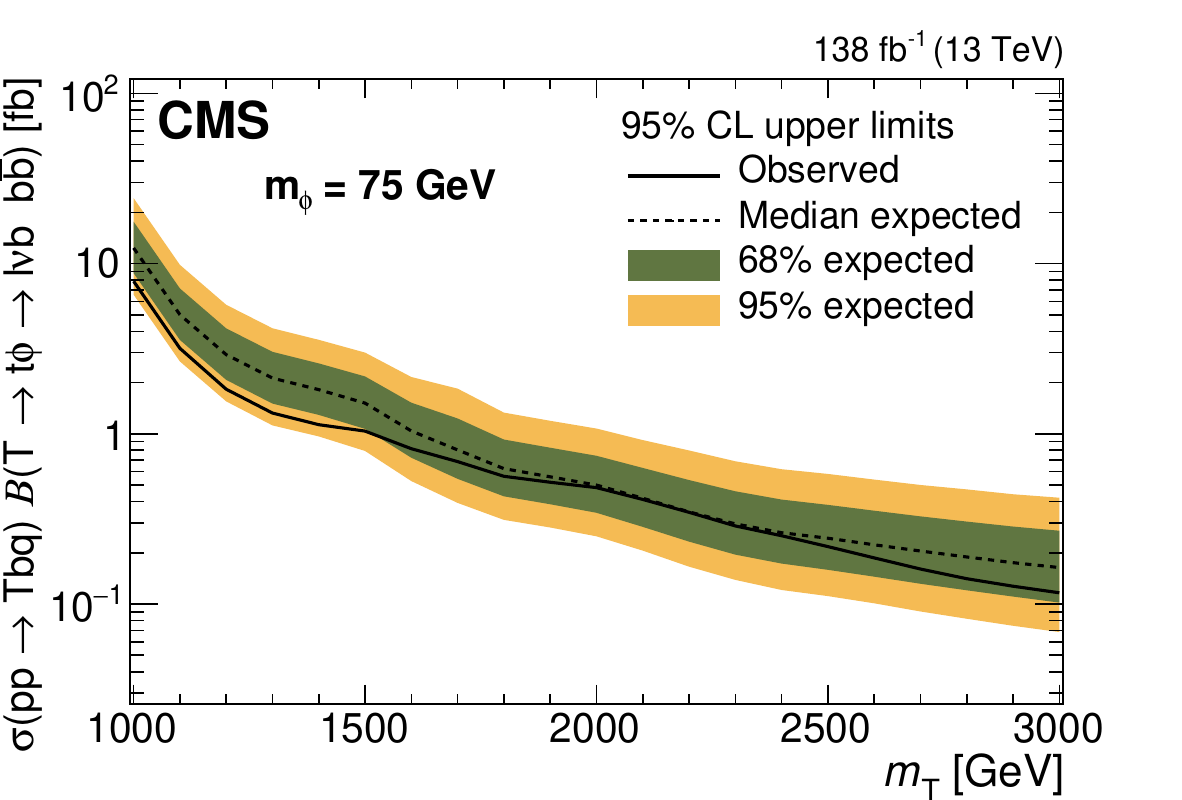}
    \includegraphics[width=0.48\textwidth]{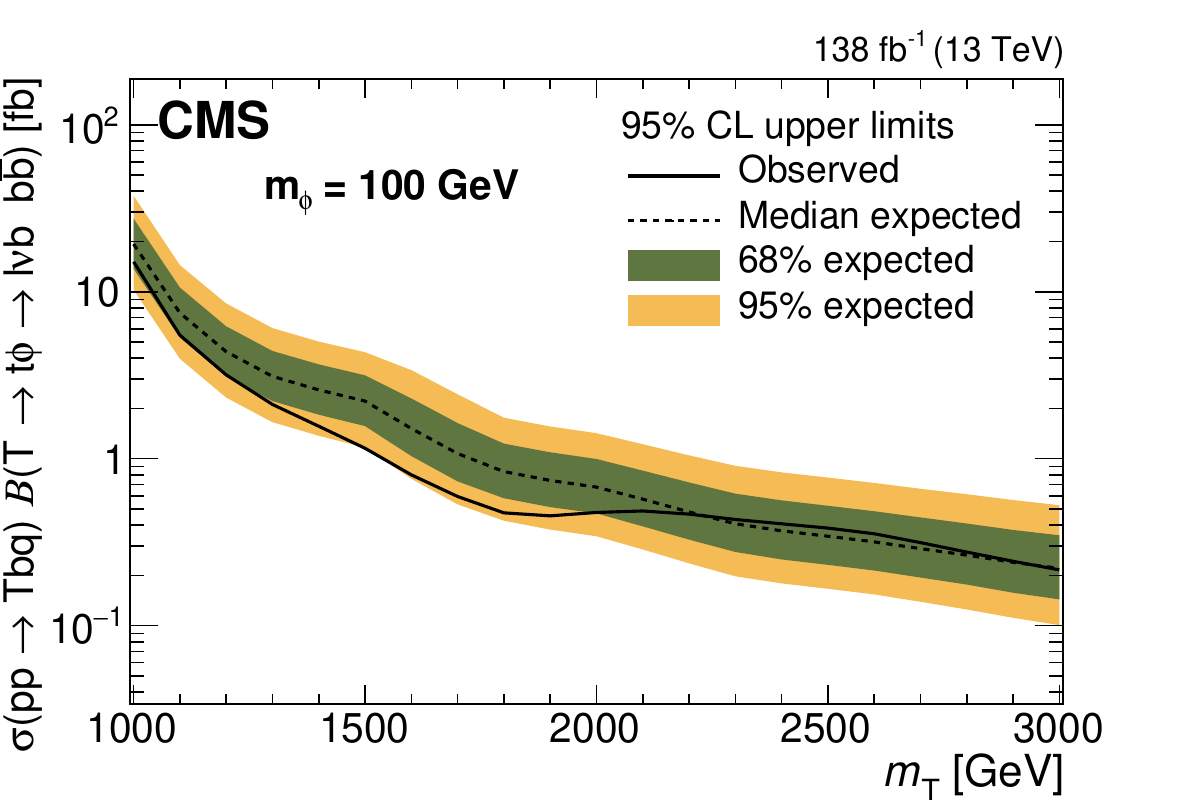}
    \caption{Observed (solid lines) and expected (dashed lines) 95\% \CL upper limits on the single \PQT quark product of cross section and
      branching ratio for the $\PQT\to\PQt\PGf \to \PQb\Pell\PGn \bbbar$ channel as a function of 
      $m_{\PQT}$, for fixed values of $m_{\PGf}$, from 25
      up to 100\GeV. The inner (green) and outer (yellow) bands represent the regions containing 68 and 95\%, respectively, of the limit
      values expected under the background-only hypothesis.}
    \label{fig:ta_brazil_1}
\end{figure}

\begin{figure}[ht!]
    \centering
    \includegraphics[width=0.48\textwidth]{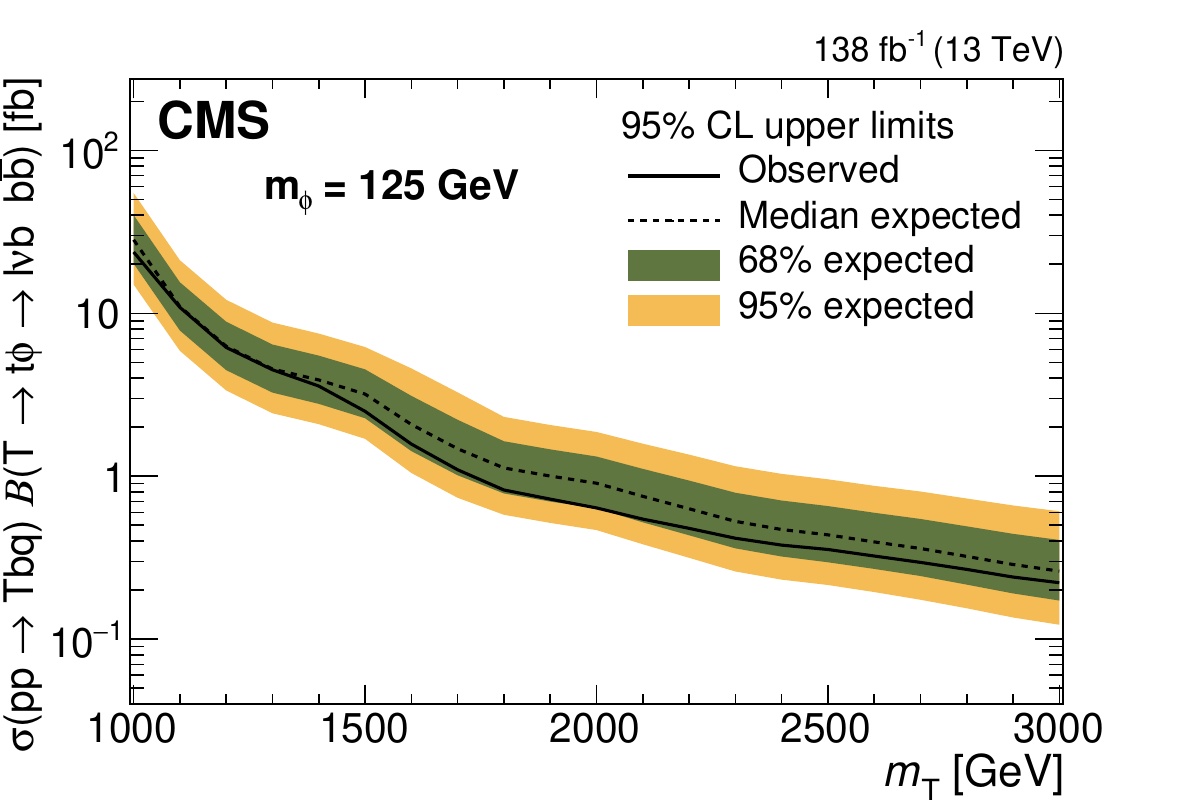}
    \includegraphics[width=0.48\textwidth]{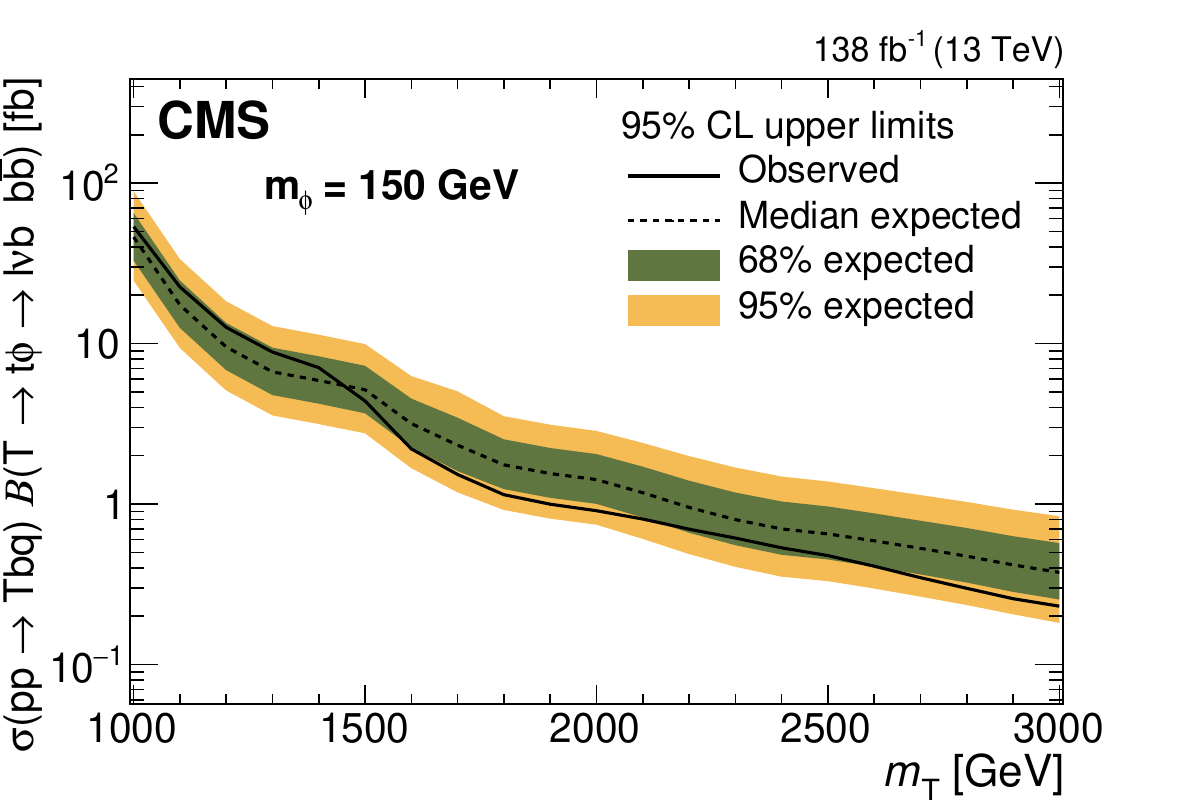} \\
    \includegraphics[width=0.48\textwidth]{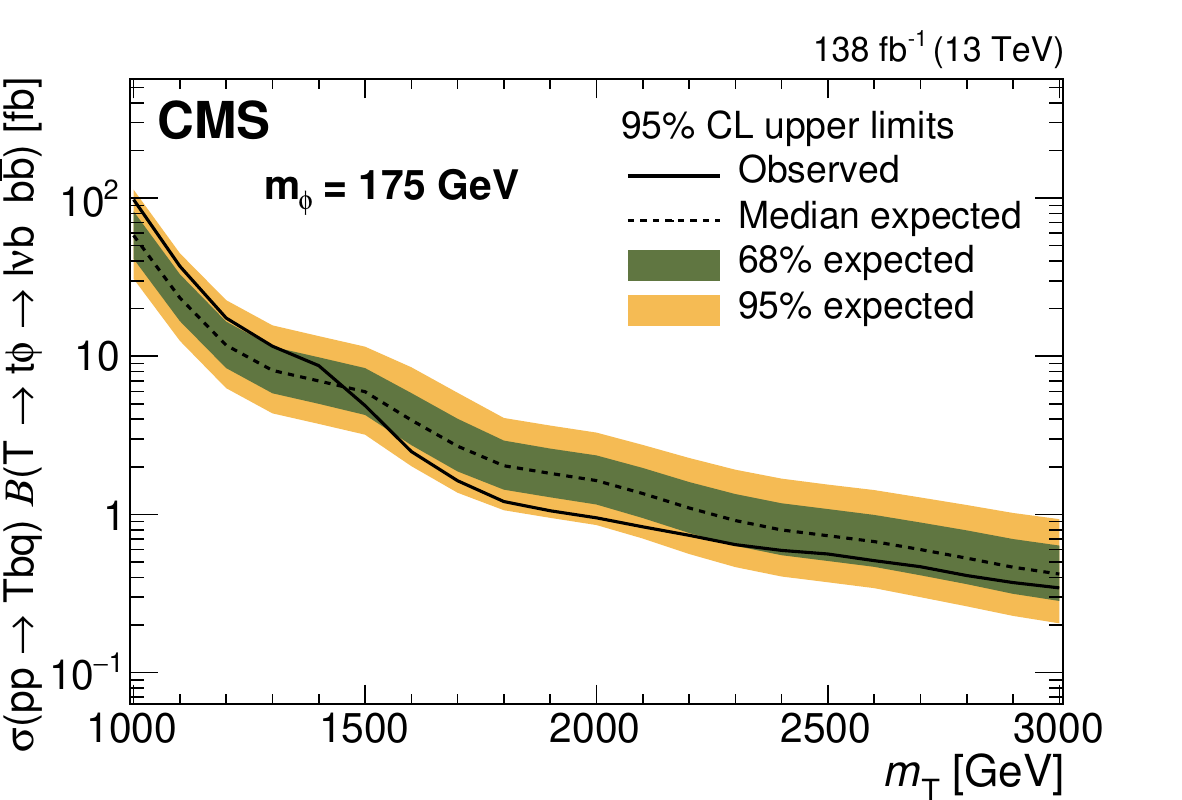}
    \includegraphics[width=0.48\textwidth]{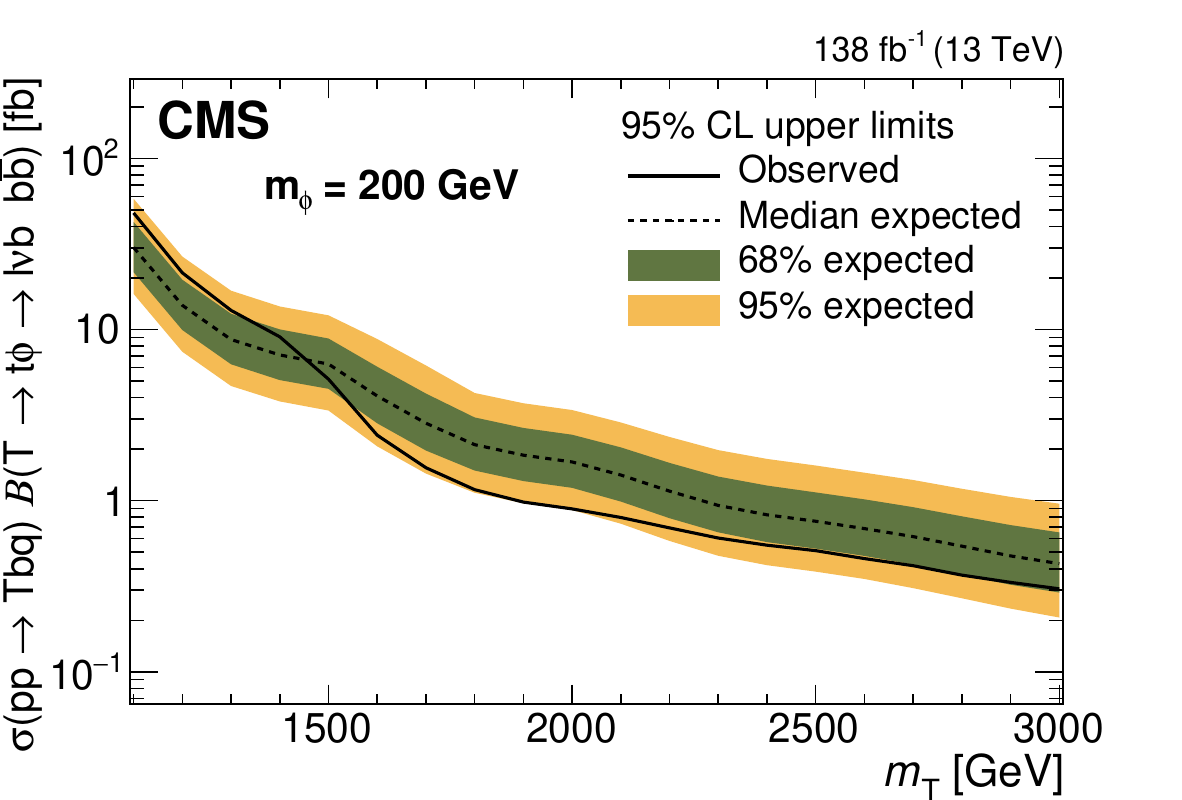} \\
    
    \includegraphics[width=0.48\textwidth]{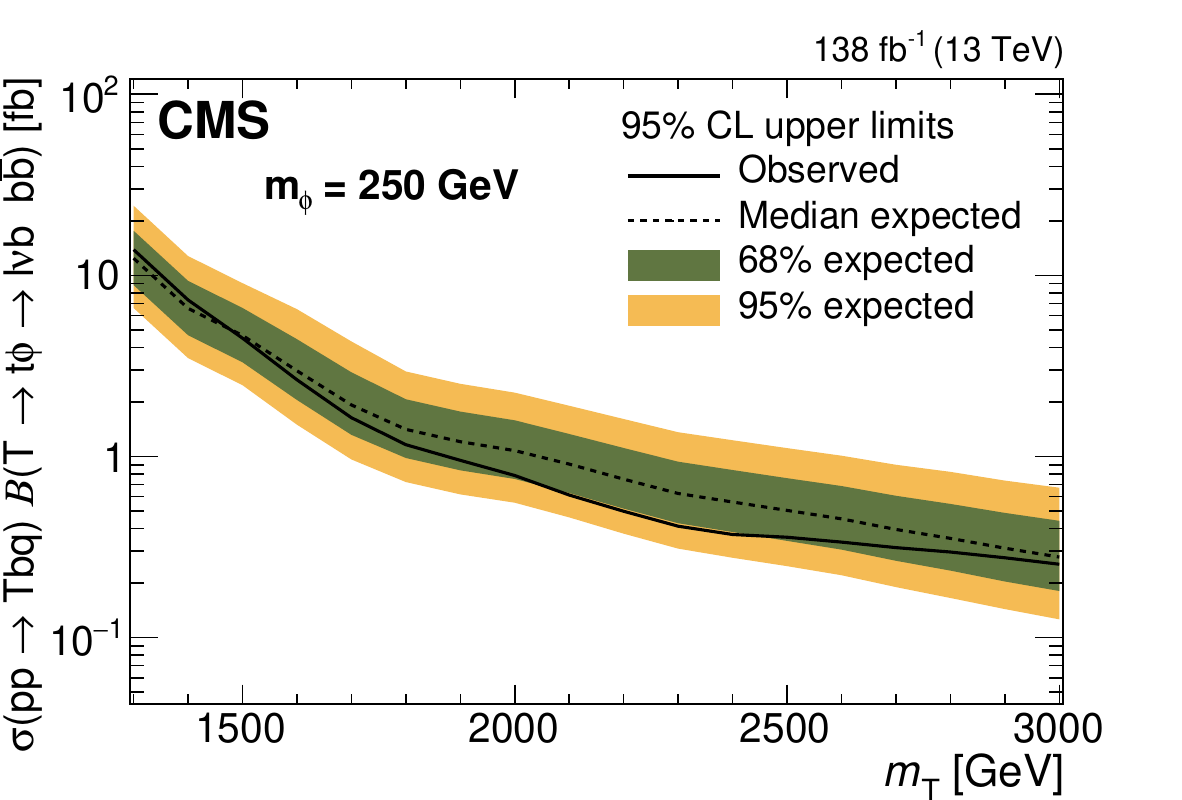}
    
    \caption{Observed (solid lines) and expected (dashed lines) 95\% \CL upper limits on the single \PQT quark product of cross section and 
      branching ratio for the $\PQT\to\PQt\PGf \to \PQb\Pell\PGn \bbbar$ channel as a function of 
      $m_{\PQT}$, for fixed values of $m_{\PGf}$, from 125
      up to 250\GeV. The inner (green) and outer (yellow) bands represent the regions containing 68 and 95\%, respectively, of the limit
      values expected under the background-only hypothesis.}
    \label{fig:ta_brazil_2}
\end{figure}

Figures~\ref{fig:ta_brazil_1} and~\ref{fig:ta_brazil_2} show the 95\% \CL upper limits
on the single \PQT quark product of cross section and 
$\PQT \to \PQt\PGf \to \PQb\Pell\PGn\bbbar$ branching fraction
as a function of $m_{\PQT}$ for fixed values of $m_{\PGf}$ in the range $[25, 250]\GeV$.
For $m_{\PGf} = 25$ (250)\GeV, values greater than 2.3 to 0.1 (14.7 to 0.3)\unit{fb}
are excluded at 95\% \CL for $m_{\PQT}$ between 1.0 and 3.0 (1.3 and 3.0)\TeV.


\begin{figure}[ht!]
\centering
\includegraphics[width=0.65\textwidth]{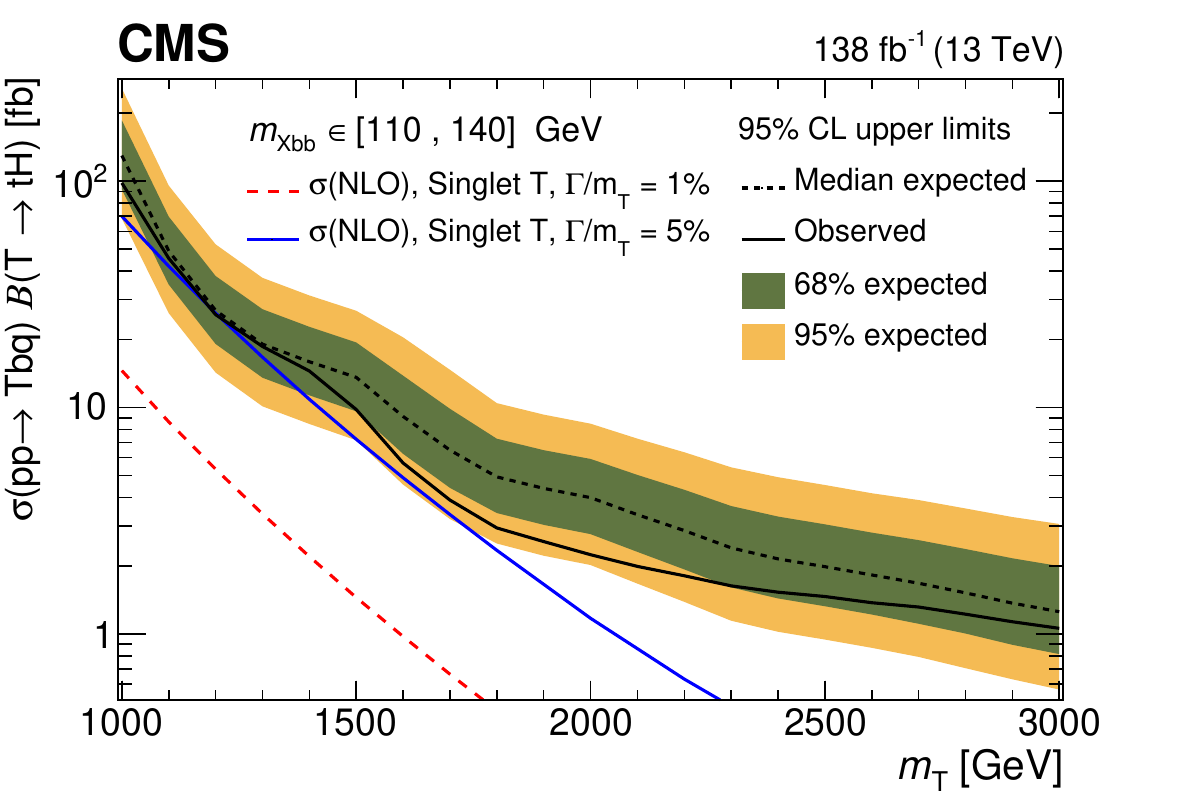}
\caption{Observed (solid black line) and expected (dashed line) 95\% \CL upper limit on the single \PQT quark production cross section times
  branching fraction for the $\PQT\to\PQt\PH$ channel as a function of the $m_{\PQT}$ mass. The inner (green) and outer (yellow) bands represent the region
  containing 68 and 95\%, respectively, of the limit values expected under the background-only hypothesis. The solid blue (dashed red) curve shows the theoretical expectation
  at NLO for a singlet \PQT quark assuming a narrow resonance with width 5\% (1\%) of the resonance mass~\cite{Buchkremer:2013bha, Carvalho:2018}.}
\label{fig:tH_brazil}
\end{figure}

In the case of the $\PQT\to\PQt\PH$ decay analysis, the 95\% \CL upper limits
on the product of the production cross section and $\PQT\to\PQt\PH$ branching fraction
as a function of $m_{\PQT}$ are shown in Fig.~\ref{fig:tH_brazil}.
Values greater than 100 to 1.0\unit{fb}
are excluded at 95\% \CL for $m_{\PQT}$ between 1 and 3\TeV.
The limits are compared with theoretical expectations assuming a singlet \PQT quark
with a resonance width $\Gamma$ of 5 or 1\% of the mass.
As a result of a reconstruction strategy optimized for large Lorentz boost of the \PQT quark decay products, the analysis is particularly sensitive to signal production for large $m_{\PQT}$ values. In the singlet scenario, this analysis improves on all previous searches performed in CMS~\cite{Sirunyan:2019xeh,CMS:2022yxp,CMS:2023agg,CMS:2024qdd,CMS:2024bni} for $m_{\PQT}$ values of 1.3\TeV and above, and it provides the strongest limits to date for $m_{\PQT}>2\TeV$. Nevertheless, the analysis sensitivity is also comparable with other searches in the range $1.1 < m_{\PQT} < 1.3\TeV$ and, for a $\Gamma/m_{\PQT}$ ratio of 5\%, it is near the level of excluding $m_{\PQT}$ around 1.2\TeV.
As a comparison, the \PQT quark search exploiting final states with one lepton and multiple jets reported by the ATLAS Collaboration~\cite{ATLAS:2023pja}, and targeting the $\PQt\PH$ and $\PQt\PZ$ decay channels, excluded \PQT quark masses, in the singlet scenario and for a $\Gamma/m_{\PQT}$ ratio of 5\%, from 1 to almost 1.2\TeV at 95\% \CL.

\clearpage

\section{Summary}\label{sec:summary}
A search for the single production of a vector-like quark \PQT with charge $2e/3$ decaying to a
top quark and a neutral scalar boson, which can be a standard model Higgs boson (\PH) or
a new scalar boson ($\PGf$), has been presented.
In the first case, a branching fraction of 25\% is assumed for the decay $\PQT\to\PQt \PH$,
while in the second case the \PQT quark is assumed to decay exclusively to $\PQt \PGf$.
Final states where the top quark decays in the lepton+jets channel and the neutral
scalar boson decays into a bottom quark-antiquark pair are considered.
The analysis is based on LHC proton-proton collision data,
corresponding to an integrated luminosity of 138\fbinv.
No significant deviation is observed between the data and the expected background. 
Upper limits at 95\% confidence level (\CL) are set on the product of the production cross
section and $\PQT \to \PQt\PGf \to \PQb\Pell\PGn\bbbar$ branching fraction
as functions of the masses of \PQT quark ($m_{\PQT}$) and of $\PGf$ boson ($m_{\PGf}$),
and assuming a \PQT quark in a narrow-width approximation produced in association
with a bottom quark.
For an $m_{\PGf}$ of 25 (250)\GeV, values greater than 2.3 to 0.1 (14.7 to 0.3)\unit{fb}
are excluded at 95\% \CL for $m_{\PQT}$ between 1.0 and 3.0 (1.3 and 3.0)\TeV.

These are the first exclusion limits set on the production
of a single \PQT quark decaying into a top quark
and a new neutral scalar boson.
The case of a vector-like quark \PQT decaying to a top quark
and a standard model Higgs boson has been studied
as well, and upper limits at 95\% \CL are set on the product of the production cross
section and $\PQT\to\PQt\PH$ branching fraction: values greater than 100 to 1.0\unit{fb}
are excluded at 95\% \CL for $m_{\PQT}$ between 1.0 and 3.0\TeV.
These results provide the best exclusion limits to date for $m_{\PQT}>2\TeV$.

\begin{acknowledgments}
We congratulate our colleagues in the CERN accelerator departments for the excellent performance of the LHC and thank the technical and administrative staffs at CERN and at other CMS institutes for their contributions to the success of the CMS effort. In addition, we gratefully acknowledge the computing centers and personnel of the Worldwide LHC Computing Grid and other centers for delivering so effectively the computing infrastructure essential to our analyses. Finally, we acknowledge the enduring support for the construction and operation of the LHC, the CMS detector, and the supporting computing infrastructure provided by the following funding agencies: SC (Armenia), BMBWF and FWF (Austria); FNRS and FWO (Belgium); CNPq, CAPES, FAPERJ, FAPERGS, and FAPESP (Brazil); MES and BNSF (Bulgaria); CERN; CAS, MoST, and NSFC (China); MINCIENCIAS (Colombia); MSES and CSF (Croatia); RIF (Cyprus); SENESCYT (Ecuador); ERC PRG, TARISTU24-TK10 and MoER TK202 (Estonia); Academy of Finland, MEC, and HIP (Finland); CEA and CNRS/IN2P3 (France); SRNSF (Georgia); BMFTR, DFG, and HGF (Germany); GSRI (Greece); NKFIH (Hungary); DAE and DST (India); IPM (Iran); SFI (Ireland); INFN (Italy); MSIT and NRF (Republic of Korea); MES (Latvia); LMTLT (Lithuania); MOE and UM (Malaysia); BUAP, CINVESTAV, CONACYT, LNS, SEP, and UASLP-FAI (Mexico); MOS (Montenegro); MBIE (New Zealand); PAEC (Pakistan); MES, NSC, and NAWA (Poland); FCT (Portugal);MESTD (Serbia); MICIU/AEI and PCTI (Spain); MOSTR (Sri Lanka); Swiss Funding Agencies (Switzerland); MST (Taipei); MHESI (Thailand); TUBITAK and TENMAK (T\"{u}rkiye); NASU (Ukraine); STFC (United Kingdom); DOE and NSF (USA).

\hyphenation{Rachada-pisek} Individuals have received support from the Marie-Curie program and the European Research Council and Horizon 2020 Grant, contract Nos.\ 675440, 724704, 752730, 758316, 765710, 824093, 101115353, 101002207, 101001205, and COST Action CA16108 (European Union); the Leventis Foundation; the Alfred P.\ Sloan Foundation; the Alexander von Humboldt Foundation; the Science Committee, project no. 22rl-037 (Armenia); the Fonds pour la Formation \`a la Recherche dans l'Industrie et dans l'Agriculture (FRIA) and Fonds voor Wetenschappelijk Onderzoek contract No. 1228724N (Belgium); the Beijing Municipal Science \& Technology Commission, No. Z191100007219010, the Fundamental Research Funds for the Central Universities, the Ministry of Science and Technology of China under Grant No. 2023YFA1605804, the Natural Science Foundation of China under Grant No. 12061141002, 12535004, and USTC Research Funds of the Double First-Class Initiative No.\ YD2030002017 (China); the Ministry of Education, Youth and Sports (MEYS) of the Czech Republic; the Shota Rustaveli National Science Foundation, grant FR-22-985 (Georgia); the Deutsche Forschungsgemeinschaft (DFG), among others, under Germany's Excellence Strategy -- EXC 2121 ``Quantum Universe" -- 390833306, and under project number 400140256 - GRK2497; the Hellenic Foundation for Research and Innovation (HFRI), Project Number 2288 (Greece); the Hungarian Academy of Sciences, the New National Excellence Program - \'UNKP, the NKFIH research grants K 131991, K 133046, K 138136, K 143460, K 143477, K 146913, K 146914, K 147048, 2020-2.2.1-ED-2021-00181, TKP2021-NKTA-64, and 2021-4.1.2-NEMZ\_KI-2024-00036 (Hungary); the Council of Science and Industrial Research, India; ICSC -- National Research Center for High Performance Computing, Big Data and Quantum Computing, FAIR -- Future Artificial Intelligence Research, and CUP I53D23001070006 (Mission 4 Component 1), funded by the NextGenerationEU program (Italy); the Latvian Council of Science; the Ministry of Education and Science, project no. 2022/WK/14, and the National Science Center, contracts Opus 2021/41/B/ST2/01369, 2021/43/B/ST2/01552, 2023/49/B/ST2/03273, and the NAWA contract BPN/PPO/2021/1/00011 (Poland); the Funda\c{c}\~ao para a Ci\^encia e a Tecnologia, grant CEECIND/01334/2018 (Portugal); the National Priorities Research Program by Qatar National Research Fund; MICIU/AEI/10.13039/501100011033, ERDF/EU, "European Union NextGenerationEU/PRTR", and Programa Severo Ochoa del Principado de Asturias (Spain); the Chulalongkorn Academic into Its 2nd Century Project Advancement Project, the National Science, Research and Innovation Fund program IND\_FF\_68\_369\_2300\_097, and the Program Management Unit for Human Resources \& Institutional Development, Research and Innovation, grant B39G680009 (Thailand); the Kavli Foundation; the Nvidia Corporation; the SuperMicro Corporation; the Welch Foundation, contract C-1845; and the Weston Havens Foundation (USA).
\end{acknowledgments}

\bibliography{auto_generated}
\cleardoublepage \appendix\section{The CMS Collaboration \label{app:collab}}\begin{sloppypar}\hyphenpenalty=5000\widowpenalty=500\clubpenalty=5000\cmsinstitute{Yerevan Physics Institute, Yerevan, Armenia}
{\tolerance=6000
A.~Hayrapetyan, V.~Makarenko\cmsorcid{0000-0002-8406-8605}, A.~Tumasyan\cmsAuthorMark{1}\cmsorcid{0009-0000-0684-6742}
\par}
\cmsinstitute{Institut f\"{u}r Hochenergiephysik, Vienna, Austria}
{\tolerance=6000
W.~Adam\cmsorcid{0000-0001-9099-4341}, J.W.~Andrejkovic, L.~Benato\cmsorcid{0000-0001-5135-7489}, T.~Bergauer\cmsorcid{0000-0002-5786-0293}, M.~Dragicevic\cmsorcid{0000-0003-1967-6783}, C.~Giordano, P.S.~Hussain\cmsorcid{0000-0002-4825-5278}, M.~Jeitler\cmsAuthorMark{2}\cmsorcid{0000-0002-5141-9560}, N.~Krammer\cmsorcid{0000-0002-0548-0985}, A.~Li\cmsorcid{0000-0002-4547-116X}, D.~Liko\cmsorcid{0000-0002-3380-473X}, M.~Matthewman, I.~Mikulec\cmsorcid{0000-0003-0385-2746}, J.~Schieck\cmsAuthorMark{2}\cmsorcid{0000-0002-1058-8093}, R.~Sch\"{o}fbeck\cmsAuthorMark{2}\cmsorcid{0000-0002-2332-8784}, D.~Schwarz\cmsorcid{0000-0002-3821-7331}, M.~Shooshtari\cmsorcid{0009-0004-8882-4887}, M.~Sonawane\cmsorcid{0000-0003-0510-7010}, W.~Waltenberger\cmsorcid{0000-0002-6215-7228}, C.-E.~Wulz\cmsAuthorMark{2}\cmsorcid{0000-0001-9226-5812}
% ...
\par}
\vskip\cmsinstskip
\dag:~Deceased\\
$^{1}$Also at Yerevan State University, Yerevan, Armenia\\
$^{2}$Also at TU Wien, Vienna, Austria\\
% ...
\end{sloppypar}
\end{document}